\documentclass[12pt]{article}

\pdfoutput=1

\usepackage{a4wide}
\usepackage{amssymb}
\usepackage{amsmath}

\usepackage{scrextend}
\addtokomafont{labelinglabel}{\sffamily}

 \numberwithin{equation}{section}

\usepackage{amsfonts}

\usepackage{epsfig}

\newcommand{\be}{\begin{equation}}
\newcommand{\ee}{\end{equation}}
\newcommand{\bea}{\begin{eqnarray}}
\newcommand{\eea}{\end{eqnarray}}

\begin{document}

\setcounter{table}{0}

\begin{flushright}\footnotesize


\end{flushright}

\mbox{}
\vspace{0truecm}
\linespread{1.1}

\vspace{0.5truecm}

\centerline{\Large \bf A dilaton-axion model for string cosmology}

\vspace{1.3truecm}

\centerline{
    {\large \bf J. G. Russo${}^{a,b}$}
   {\bf and} 
   {\large \bf P. K. Townsend${}^{c}$}
 }

\vspace{0.8cm}

\begin{labeling}{u}

\item [${}^a$]{{\it Instituci\'o Catalana de Recerca i Estudis Avan\c{c}ats (ICREA),\\
Pg. Lluis Companys, 23, 08010 Barcelona, Spain.}}

\item [${}^b$]{{\it  Departament de F\' \i sica Cu\' antica i Astrof\'\i sica and Institut de Ci\`encies del Cosmos,\\ 
Universitat de Barcelona, Mart\'i Franqu\`es, 1, 08028
Barcelona, Spain. }}

\item [${}^c$]
{{\it Department of Applied
Mathematics and Theoretical Physics,\\ 
Centre for Mathematical
Sciences, University of Cambridge,\\
Wilberforce Road, Cambridge, CB3
0WA, UK.  }} 

\end{labeling}

\noindent {\it E-Mail:}  {\texttt jorge.russo@icrea.cat, pkt10@cam.ac.uk} 

\vspace{1.2cm}

\centerline{\bf ABSTRACT}
\medskip

The generic scale-invariant theory of an axion and a dilaton coupled to gravity in $d$-dimensions is generalized to 
a `universal' one-axion model with two dilatons that reproduces itself under consistent dimensional-reduction/truncation.
Flat FLRW cosmologies are shown to correspond to trajectories of a three-dimensional autonomous dynamical system, 
which we analyse with a focus on  accelerated  cosmic expansion, deriving the precise swampland bounds that exclude eternal acceleration. We also show that for two sets of values of its three 
independent parameters, the model is a consistent truncation of  maximal `massive' supergravity theories 
arising from string/M-theory; for these maximal-supergravity parameter values the FLRW cosmologies include some with a transient de Sitter-like phase, but not the recurring de Sitter-like phase or eternal cosmic acceleration that is 
possible for other parameter values.

\noindent

\newpage

\tableofcontents

\section{Introduction: the model}
\setcounter{equation}{0}

The current $\Lambda$CDM model of cosmology assumes a small positive cosmological constant. As this implies
a late-time de Sitter geometry, an interpretation of the model as an effective theory obtainable from a 
compactification of string/M-theory requires the existence of some compactification  to a four-dimensional de Sitter universe. 
So far, no compelling ``de Sitter compactification'' has been found, but the observational evidence for accelerated cosmic 
expansion could still be explained by a string compactification to some other cosmological spacetime that approximates 
de Sitter for some sufficiently long period. This is {\it a priori} plausible because string-M-theory requires 
us to interpret any cosmological constant in the effective four-dimensional theory as the expectation value of some scalar field, and it would be naive to suppose that scalar expectation values will remain constant. In this context, the absence of a compactification to de Sitter spacetime indicates only that the potential energy function $V$ for scalar fields in the effective four-dimensional gravitational theory has no stationary points in the region for which $V>0$. 

The absence of a stationary point of $V$ with $V>0$ still allows ``transient cosmic acceleration'' in the context of generic homogeneous and isotropic (FLRW) cosmologies, and there exist string/M-theory compactifications to such cosmological spacetimes (e.g. \cite{Townsend:2003fx,Ohta:2003pu}). For models with a single scalar field, this can be understood from the fact that there must be some cosmological time $t_*$ on any FLRW cosmological trajectory in a region with $V>0$ at which $V$ reaches a maximum value, at which time the kinetic energy will be zero, implying an equation of state that temporarily supports an FLRW cosmology with constant Hubble parameter \cite{Emparan:2003gg}; in this sense the FLRW spacetime approximates de Sitter space in some time interval around $t=t_*$. With more than one scalar field, the kinetic energy need not be zero at $t=t_*$ but one may always choose initial conditions for which the potential energy dominates, leading to a transient de Sitter-like phase, although this may require some fine tuning. However, the main difficulty is that the duration of the period of transient acceleration is too short to be useful in known examples of cosmological compactifications of the 10/11-dimensional supergravity theories that serve as effective theories for string/M-theory \cite{Ohta:2003ie,Wohlfarth:2003kw}. 

It was pointed out in \cite{Sonner:2006yn} that a simple dilaton-axion model, with an exponential potential for the dilaton,  
typically has cosmological solutions describing flat  expanding FLRW (homogeneous and isotropic) universes that undergo 
recurring periods of cosmic acceleration before finally approaching a late-time scaling solution. As we pointed out in \cite{Russo:2018akp}, this is because the kinetic energy cycles between the dilaton field and the axion field.
This phenomenon has the potential  to lengthen periods of 
transient cosmic acceleration, as well as making repetitions of these periods possible, but no attempt was made in \cite{Sonner:2006yn,Russo:2018akp} to investigate this possibility in more complicated models or to make 
contact with string/M-theory.  

Motivated by supergravity/string-theory considerations that we summarize below, we consider here a generalisation of the model considered in \cite{Sonner:2006yn} to one with one axion and {\it two} dilatons, in a $d$-dimensional spacetime. The Lagrangian density is 
\be\label{basicmodel}
{\mathcal L} = \frac12 \sqrt{-\det g_{(d)}} \left\{2R - (\partial\phi_1)^2 - (\partial\phi_2)^2 - e^{\vec\mu\cdot \vec \phi} (\partial\chi)^2  - 
m^2 e^{\vec\lambda\cdot \vec\phi}\right\}\, , 
\ee
where $R$ is the scalar curvature for the spacetime $d$-metric $g_{(d)}$, and $\vec\phi= (\phi_1,\phi_2)$ are the 
two dilaton fields. The one axion field is $\chi$; by `axion' we mean a (pseudo)scalar field that appears in the action only through its derivatives. The scalar-field interactions are determined by the pair of 
2-vector coupling constants $(\vec\mu, \vec\lambda)$ up to equivalence under $O(2)$-transformations, since the dilaton kinetic term is $O(2)$ invariant. We may use this equivalence to choose 
\be\label{O2frame}
\vec\lambda= (\lambda_1,\lambda_2) \, , \qquad \vec\mu = (\mu,0) \, , \quad \mu\ge0\, . 
\ee
We shall find it convenient to define new polar parameters $(\lambda,\theta)$ by 
\be\label{polar}
\lambda_1+ i\lambda_2 = \lambda e^{i\theta}\, .  
\ee
When $\sin\theta=0$, setting $\phi_2=0$ is a consistent truncation\footnote{A truncation is ``consistent'', in the sense used here, if all solutions of the truncated theory are also solutions of the untruncated theory with the to-be-truncated fields set to zero.} to the dilaton-axion model considered in \cite{Sonner:2006yn}, but the $\sin\theta=0$ case appears not to be relevant to string/M-theory, as we shall see. One possible reason for this emerges from a consideration 
of how the exponential potential term in \eqref{basicmodel} breaks the symmetries of the `massless' ($m=0$) model.

Apart from diffeomorphism invariance, the continuous symmetries for $m=0$ comprise a simple scaling symmetry (for which both the metric and scalars have 
scaling dimension zero) and an invariance under the $Sl(2;\mathbb{R})\times \mathbb{R}$ isometry group of the scalar field target space; the  $Sl(2;\mathbb{R})$ factor acts by fractional linear transformations on $(\phi_1,\chi)$ 
(which parametrize a hyperbolic 2-space with inverse radius of curvature $\mu/2$) while the $\mathbb{R}$ factor acts by 
constant shifts of $\phi_2$. The presence of the potential term for $m\ne0$ preserves the combination of the scaling symmetry with an $\mathbb{R}$ subgroup of the target space isometry group, but this subgroup depends on whether $\sin\theta$ is zero or non-zero, in the following way:
\begin{itemize}
    \item $\sin\theta\ne 0$. In this case the scalar potential preserves a combination of the scaling 
    symmetry with the $\mathbb{R}$ factor of the isometry group (which acts by shifting $\phi_2$). It also preserves
    the one-dimensional subgroup of $Sl(2;\mathbb{R})$ that acts by shifting $\phi_1$, although the entire 
    $Sl(2;\mathbb{R})$ symmetry group is preserved when $\cos\theta=0$. 
    
    \item $\sin\theta=0$. In this case the scalar potential term preserves a combination of the scaling symmetry with the one-dimensional subgroup of $Sl(2;\mathbb{R})$ that acts by a scaling of $\chi$ (which therefore acquires a non-zero scaling dimension) combined with a shift of $\phi_1$. It also preserves the $\mathbb{R}$ factor of the isometry group that acts by a constant shift of $\phi_2$.
    \end{itemize}
  This discontinuity in the pattern of symmetry breaking that occurs as $\sin\theta\to0$ suggests that the one-axion model defined by 
  \eqref{basicmodel} for $\sin\theta\ne 0$ will have  different features for $\sin\theta=0$ and $\sin\theta\ne0$. 
  
  We have already seen that a special feature of $\sin\theta=0$ is the possibility of a consistent truncation to the model of \cite{Sonner:2006yn}. 
A significant special feature of the one-axion model defined by \eqref{basicmodel} with  $\sin\theta\ne0$ is that it reproduces itself under consistent dimensional reduction/truncation. Specifically, $S^1$-reduction from $d\ge4$ to $(d-1)$ dimensions yields a similar model with an additional dilaton $\phi_3$ (and 
a Kaluza-Klein 1-form gauge field that may be consistently set to zero); one obtains 
\be
{\mathcal L} \to  \frac12 \sqrt{-\det g_{(d-1)}} \left\{2R - (\partial\phi_1)^2 - (\partial\phi_2)^2 
- (\partial\phi_3)^2 -  e^{\boldsymbol{\mu} \cdot \boldsymbol{\phi}} (\partial\chi)^2  
- m^2 e^{\boldsymbol{\lambda}\cdot \boldsymbol{\phi}}
\right\}\, , 
\ee
where $g_{(d-1)}$ is the (Einstein-frame) $(d-1)$-metric, and $\boldsymbol{\phi} = (\phi_1,\phi_2,\phi_3)$; the 
new 3-vector parameters are
\be
\boldsymbol{\mu} = (\mu,0,0)\, ,   \qquad \boldsymbol{\lambda} = (\lambda_1,\lambda_2, \gamma) \, , \qquad \gamma = \sqrt{\frac{2}{(d-2)(d-3)}}\, . 
\ee
We may now consistently truncate to a two-dilaton model by setting 
\be
\lambda_2\phi_3 -  \gamma \phi_2 =0\, . 
\ee
For $\lambda_2=0$ this implies $\phi_2=0$ but we can then rename $\phi_3$ as $\phi_2$. For $\lambda_2\ne0$  we can eliminate $\phi_3$ and then rescale $\phi_2$ to achieve a canonical normalization for it. In either case, we arrive at a Lagrangian density of the form (\ref{basicmodel}) that we started with,  but with $\lambda_2 \to \sqrt{\lambda_2+ \gamma^2}$. As both $\mu$ and $\lambda_1$ are unchanged we deduce that both $|\vec\mu|$ and $\vec\mu\cdot\vec\lambda$ are $d$-independent, whereas $\lambda$ is $d$-dependent; in agreement with \cite{Lu:1995hm}, we find that this dependence is such that 
\be
\label{laddd}
\lambda^2 = \Delta + \frac{2(d-1)}{d-2} \, ,  
\ee
for some $d$-independent number $\Delta$. Furthermore, as 
\be
\lambda\cos\theta = \frac{\vec\mu\cdot\vec\lambda}{|\vec\mu|}\, , 
\ee
which is independent of $d$, the $d$-dependence of $\lambda$ implies a corresponding $d$-dependence of $\theta$. 
In particular, if $\theta=0$  initially, it will not remain zero as we dimensional reduce and then consistently truncate to another one-axion model, which is another way of saying that two dilatons are required for a one-axion model to reproduce itself (up to changes in the parameters) in this process of dimensional-reduction/truncation.  

Because of this self-reproducing feature we shall refer to the model  defined by \eqref{basicmodel}, as the ``universal one-axion model'', it being  understood that we allow for all non-zero 2-vector parameters $(\vec \lambda,\vec \mu)$.  For some purposes, it is more convenient to exchange the three independent parameters $(\mu,\lambda; \theta)$ for three independent parameters that are also independent of both $d$ and the choice of $O(2)$ basis for the dilatons; we choose
\be\label{defX}
\mu = |\vec\mu| \, , \qquad \Delta\, , \qquad X= \frac{\vec\mu\cdot\vec\lambda}{\mu^2}  \, . 
\ee
All essential aspects of our results will depend only on these
parameters and are therefore dimension independent.

One purpose of this paper is to study the FLRW cosmological solutions of the universal one-axion model. We focus on the solutions that describe a flat expanding universe, partly because these are late-time attractors for accelerated expansion but also because this allows us to reduce the problem to one of finding solutions of a three-dimensional autonomous dynamical system, which generalises the two-dimensional system analysed in \cite{Sonner:2006yn}. The step from a two-dimensional to a three-dimensional system is potentially significant since ``strange'' attractors are possible in the three-dimensional case, and are a common feature of various 3-dimensional autonomous systems that are superficially similar to the one that we find. However, the dynamics of our three-dimensional  system turns out to be not so different from the two-dimensional one of \cite{Sonner:2006yn} because one variable  either increases or decreases monotonically on all trajectories. This implies, for example, that there are no strange attractors and no chaotic behaviour. 

Fixed points of the dynamical system correspond to scaling solutions
of the Einstein-scalar equations associated with a particular equation of state specified by the pressure to density ratio $w$
with $|w|\le 1$; its value determines whether the cosmic expansion is decelerating or accelerating; $w=-1$ corresponds to de Sitter space and we shall say that the cosmology is ``de Sitter-like" if $w=-1+\delta$ for sufficiently small $\delta$. On other (non-fixed point) trajectories the parameter $w$ varies continuously, and
trajectories on which $w$ approaches $-1$ will have a transient de Sitter-like phase. 

The difficulty of finding string compactification to de Sitter space
has been formalized in recent years in terms of postulated swampland conditions on the scalar field potential in effective gravity-scalar theories. The underlying reason for the swampland conditions is unknown but a plausible hypothesis is that they arise from a
principle of no eternal cosmic acceleration, which is needed to avoid a future cosmological horizon; these horizons present obstacles to the construction of any unitary theory of quantum gravity, string/M-theory in particular \cite{Witten:2001kn,Dyson:2002pf,Banks:2002wr}. Here we use this hypothesis 
to derive precise swampland bounds for the universal one-axion model; it turns out that  a constraint on the gradient of the potential (of the kind proposed in \cite{Obied:2018sgi}) is insufficient; another constraint involving the curvature of the metric on the scalar-field target space is also needed. 

Another purpose of this paper is to apply the results of our cosmological investigation of the universal one-axion model to the  $d\le8$ maximal supergravity theories that are effective field theories for supersymmetry-preserving reductions of string/M-theory;
as we need a positive potential term, this means that we will be considering `massive' versions of the $d\le8$ maximal supergravity theories\footnote{We stress that ``massive'' refers here to the presence of a positive potential for one or more scalar fields in the graviton supermultiplet, in contrast to the more recent usage of ``massive''  for theories with a massive graviton.}. Some general aspects of cosmological solutions of these massive supergravity theories were considered in \cite{Lu:1996er} but our focus, based on the universal one-axion model introduced above, is different. 

Many examples of massive maximal supergravity theories are known in various dimensions, e.g. \cite{Aurilia:1980xj,Romans:1985tz,Bergshoeff:1996ui, Cowdall:1996tw} for $d=4$, $d=10$, $d=9$ and $d=8,7$, respectively, in addition to general methods \cite{Cowdall:1996tw,Lavrinenko:1996mp,Kaloper:1998kr} that are believed to generate the most general massive deformation of the unique `massless' maximal supergravity in lower dimensions (as has been established for the half-maximal case \cite{Kaloper:1999yr}). Here we show that the universal one-axion model is a consistent truncation of all of the above massive maximal supergravity theories except for the few for which there is no axion field. In this maximal supersymmetry context we find that $\mu=2$ and $\Delta=4$, and that $X$ may have one of two values: $(0,\tfrac12)$. For massive supergravity theories that are not maximally supersymmetric (e.g. those found from a combination of Scherk-Schwarz  reduction with compactification on Ricci-flat manifolds such as $K_3$ or Calabi-Yau \cite{Lavrinenko:1996mp}) we expect other values of $(\mu,\Delta,X)$ to occur (since, for example,  it is known that $\Delta=2$ for the 
half-maximal case). 

The organization of this paper is as follows: we first discuss the flat FLRW cosmological solutions of the universal one-axion model defined by \eqref{basicmodel}, determining the global phase-space for the associated 3-dimensional autonomous dynamical system, and the qualitative behaviour of trajectories. We then discuss how the equation of state for the scalar-field matter evolves on trajectories, the condition for cosmic accelerated expansion at fixed points, and the circumstances in which a transient de Sitter-like phase occurs, and potentially re-occurs, and we use these results to obtain our cosmological swampland bounds. We then show how the universal one-axion model arises as a consistent truncation of various `massive' maximal supergravity theories. Finally, we summarize and discuss some further issues raised by our results.

\section{Flat FLRW cosmologies}\label{sec:FLRW}

Using the $O(2)$-frame choice \eqref{O2frame} and the notation of \eqref{polar}, the Lagrangian density for the
`universal' one-axion model defined by \eqref{basicmodel} becomes 
\be\label{modelogeneral}
{\mathcal L} = \frac12 \sqrt{-\det g_{(d)}} \left\{2R - (\partial\phi_1)^2 - (\partial\phi_2)^2 
- e^{\mu \phi_1} (\partial\chi)^2  -  m^2 e^{\lambda(\phi_1\cos\theta + \phi_2\sin\theta)} \right\}\, . 
\ee
We are interested in cosmological solutions of this model, and we focus on flat 
FLRW universes with a $d$-metric of the form 
\be\label{FLRW}
ds^2= -e^{2\alpha \varphi} f^2 d\tau^2+e^{2\beta\varphi } d\ell^2(\mathbb{E}^{d-1})\, , 
\ee
where 
\be\label{alphabeta}
\alpha = \sqrt{\frac{d-1}{2(d-2)}}\, ,\qquad   \beta = \frac{\alpha}{d-1}\ ,  
\ee
and the function $f=f(\tau)$ is smooth and strictly monotonic but otherwise arbitrary; it is included 
to allow for an arbitrary time parametrization.  The FLRW scale factor is 
\be
S(t) = e^{\beta\varphi(\tau)}\, ,  \qquad dt= e^{\alpha\varphi} f d\tau\, , 
\ee
where $t$ is the FLRW ``cosmological time'' $t$.
Additionally, we assume that all scalar fields are functions only of the time variable $\tau$.

For this metric ansatz, the Einstein and scalar field equations that follow from \eqref{modelogeneral}
become the Euler-Lagrange equations associated to the following effective Lagrangian: 
\be\label{lagzzz}
L_{\rm eff} = \frac{1}{f}\bigg(
- \dot \varphi^2+\dot \phi_1^2+\dot \phi_2^2
+ e^{\mu \phi_1}\dot \chi ^2\bigg)
-fm^2 e^{\lambda (\phi_1\cos\theta +\phi_2\sin\theta) + 2\alpha \varphi}\, ,
\ee
where the overdot indicates a derivative with respect to $\tau$. The equation for $\chi$ has the first integral
\be\label{solchi}
f^{-1}\dot \chi = J e^{-\mu \phi_1}\ ,
\ee
where $J$ is the integration constant. In principle we could use this to arrive at a set of ($J$-dependent) equations 
for $(\varphi, \vec\phi)$ alone; this is useful for some special cases but it is advantageous to proceed differently
for the general case. 

Following \cite{Sonner:2006yn}, we fix the time-reparametrization invariance by choosing
$f$ to cancel the field-dependence of the potential term in the equations of motion; this 
cancellation occurs for the choice
\be\label{tiempotau}
f= e^{-\tfrac12 \lambda (\phi_1\cos\theta +\phi_2\sin\theta) -\alpha \varphi}\, .
\ee
This implies that 
\be
dt= e^{-\tfrac12 \lambda (\phi_1\cos\theta +\phi_2\sin\theta)} \, d\tau\, ,  
\ee
which fixes the relation between our preferred time parameter $\tau$ and the FLRW cosmological time $t$. 
For this choice of time parametrization, the algebraic field equation for $f$ reduces to the constraint
\be\label{constraint}
e^{\mu  \phi_1}\dot \chi^2 = v^2 - u_1^2-u_2^2- m^2 \ , 
\ee
where
\be
v=\dot\varphi\, , \qquad u_1= \dot\phi_1\, , \qquad  u_2=\dot\phi_2\, . 
\ee
The equations of motion for $(\varphi,\ \phi_1,\ \phi_2)$, and with $f$ given by \eqref{tiempotau}, are
\bea\label{autonixx}
\dot v &=&  \alpha  m^2-\frac{1}{2} v \left(\lambda  u_1\cos\theta +\lambda u_2\sin\theta +2 \alpha  v\right)
\nonumber\\
\dot u_1 &=&  -\frac12 m^2 \lambda \cos\theta-\frac12 u_1
   \left(\lambda  u_1 \cos\theta + \lambda  u_2 \sin\theta  +2 \alpha  v\right)
   \nonumber\\
 &&\qquad +\ \frac{\mu}{2}   \left(v^2-u_1^2-u_2^2- m^2\right)  \\
\dot u_2 &=& -\frac12 m^2 \lambda \sin\theta 
-\frac{1}{2} u_2 \left(\lambda  u_1 \cos\theta + \lambda  u_2 \sin\theta 
+2 \alpha  v\right) \nonumber \, , 
\eea
where we have used the constraint to eliminate the $\dot \chi$ dependence in the equation for $\dot u_1$. The resulting three first-order differential equations for $(v,u_1,u_2)$  define a 3-dimensional autonomous dynamical system, which generalises the 2-dimensional autonomous dynamical system analysed in \cite{Sonner:2006yn}. 
Notice that the right hand sides of the three equations of \eqref{autonixx} are all quadratic functions of the three variables. A number of classic examples of three dimensional systems that exhibit chaotic 
behaviour and strange attractors, such as the Lorenz and R\"ossler systems, are of this type. 
Therefore, one important question is whether the dynamics of the system defined by \eqref{autonixx}
 can be chaotic in some parameter regime. 
 
We shall present a comprehensive analysis of
the trajectories in the following subsections, but first we need to consider the role of the Friedman constraint 
\eqref{constraint}, which restricts the variables $(v,u_1,u_2)$ to the region for which
\be\label{positivo}
V\ge 0\, , \qquad V\equiv  v^2 - u_1^2-u_2^2- m^2 \, . 
\ee
The restriction to this region is consistent because a trajectory passing through any point in it lies entirely in it. 
This follows from the fact that the surface $V=0$ is an invariant submanifold of the autonomous system. To see this we use
\eqref{autonixx} to deduce that
\be\label{reccc}
\dot V = -V U\ ,\qquad U \equiv  
2\alpha v +\lambda \sin\theta u_2+(\mu+\lambda\cos\theta)u_1 \, .  
\ee
This shows that $\dot V=0$ if $V=0$. All trajectories are therefore in one of two regions, each bounded 
by a branch of the hyperboloid $V=0$. In the region bounded by the upper branch of this hyperboloid we have $v>0$, which implies an expanding universe. The other region yields the time-reversed trajectories on which the universe is contracting, so we may restrict our analysis to the upper region. 

\subsection{Global phase-space}

The global properties of the autonomous system defined by \eqref{autonixx} are best studied in 
the following new variables:
\be
z=\frac{m}{v}\ ,\qquad x=\frac{u_1}{v}\ ,\qquad y=\frac{u_2}{v}\, , 
\ee
This is a generalization of the global coordinates used in \cite{Sonner:2006yn} (but $(x,y)$ there become 
$(x,z)$ here, and $y$ is a new variable). 
In these variables, the constraint $V\ge0$ restricts trajectories to the unit ball
\be
\tilde V \ge 0\ , \qquad \tilde V\equiv 1- x^2-y^2-z^2\, .  
\label{esfera}
\ee
The restriction that we make to $v>0$ further limits us to the upper-half ball, {\it i.e.} $z\ge0$. 
The equations \eqref{autonixx} defining the autonomous dynamical system 
in this half-ball can now be written as 
\bea\label{3Dsystem}
x' &=& -z^2\left(\hat\lambda\cos\theta + x \right)    
+\hat\mu \left( 1-x^2-y^2-z^2 \right) \nonumber \\
y' &=& - z^2 \left(\hat\lambda\sin\theta + y\right) \\
z' &=& z\left[\hat\lambda\left(x\cos\theta  + y\sin\theta\right) + (1-z^2)\right]\, , \nonumber
\eea
where the prime indicates a derivative with respect to a new time variable $\tilde\tau$, defined by 
\be 
d\tilde\tau = 2\alpha \left(\frac{m}{z}\right) d\tau \qquad (z>0),   
\ee
and
\be
\qquad \hat\mu= \frac{\mu}{2\alpha}\, , \qquad \hat\lambda= \frac{\lambda}{2\alpha}\, . 
\ee
In terms of $\hat\lambda$, the relation \eqref{laddd} to the $d$-independent parameter $\Delta$ is 
\be\label{hatlamDelta}
\hat\lambda^2 = 1 + \frac{\Delta}{(2\alpha)^2}\, . 
\ee

Now we have 
\be
\tilde V'= -2(\hat\mu x+ z^2)\tilde V \, , 
\ee
which confirms that the upper-half-sphere boundary of the upper-half-ball
is an invariant submanifold. Another invariant submanifold is $z=0$, the plane ``at infinity'', 
so trajectories that pass through an interior point of the upper-half-ball remain in it.
The dynamical subsystems on these two boundary submanifolds are as follows:
\begin{itemize}

\item $z=0$:
    \be
    x'= \hat\mu (1-x^2-y^2) \,,  \qquad y'=0\, . 
    \ee
    Notice that trajectories on this plane are lines of constant $y$ on which $x$ increases monotonically since $x^2+y^2\le1$. They start and end on the circle 
    $x^2+y^2=1$, which is a circle of fixed points. 
    
    \item $\tilde V=0$:
    \bea\label{bndry}
    x' &=& -(1-x^2-y^2)(\hat\lambda\cos\theta + x) \, , \nonumber \\
    y' &=& -(1-x^2-y^2)(\hat\lambda\sin\theta + y)\, . 
    \eea
    There is a circle of fixed points at $x^2+y^2=1$ and 
    for $\hat\lambda<1$ there is also a fixed point at 
    \be\label{bfp}
    (x,y) = -\hat\lambda(\cos\theta, \sin\theta) \qquad (\hat\lambda<1).
    \ee
    All trajectories with $x^2+y^2<1$ are solutions of 
    \be\label{bts}
    x\sin(\psi-\theta)  + y\cos(\psi-\theta) + \hat\lambda \sin\psi = 0\, , 
    \ee
    for some angular constant $\psi$. For $\hat\lambda<1$ this equation is an identity at the isolated fixed point, and 
    other solutions are lines emanating from this fixed point  that end on the circle of fixed points 
    $x^2+y^2=1$. As $\hat\lambda\to 1$ the isolated fixed point moves to coincide with a fixed point
    on the circle of fixed points  and all trajectories now start at this point and end elsewhere on the circle of fixed points. 
    As $\hat\lambda$ increases further, the trajectories start at points on an arc of the circle of fixed points and end on 
    the complementary arc. 
        
     \end{itemize}
   
A simplifying feature of the full 3-dimensional dynamical system defined by \eqref{3Dsystem} is that the the plane $y= - \hat\lambda\sin\theta$ is an invariant submanifold. However, this is relevant to the subsystem determining trajectories in the allowed upper-half ball only if this plane intersects it, which it will iff $\hat\lambda^2 \sin^2\theta <1$. Using \eqref{hatlamDelta}, we can rewrite this condition as 
\be\label{A>0} 
A>0\ ,\qquad   A :=1-\hat \lambda^2 \sin^2\theta =\frac{1}{(2\alpha)^2} \left(X^2\mu^2 - \Delta\right)\,   ,
\ee 
where $X$ is the variable defined in \eqref{defX}.  If $A\le0$ the only fixed points in the allowed region of the 
system \eqref{3Dsystem} are those on the circle $x^2+y^2=1$ in the plane $z=0$. 
The trajectories on the $z=0$ plane are lines of constant $y$. On all other trajectories (in the upper-half ball) $y$ is monotonically decreasing, so all trajectories start on the circle of fixed points at $z=0$ and end on this circle of fixed points. This is illustrated in Figs. \ref{primera}, \ref{segunda}.

 If $A>0$, the global topology of trajectories is more complicated because there can be fixed points not on the $z=0$ plane. However, all these additional fixed points are on the invariant plane $y= - \hat\lambda\sin\theta$, which separates the trajectories into two disjoint parts.  As is clear from the equations \eqref{3Dsystem}, $y$ changes monotonically in each part 
 such that all trajectories end either at fixed points in the invariant plane (which we investigate in the following subsection) or at fixed points on the  attractive arc in the $z=0$ plane. Examples are shown in figs. \ref{tercera}a,b and \ref{cuarta}a,b. 
 Note that the strict monotonicity of $y$ on trajectories not in the invariant plane excludes the possibility of chaotic behaviour, which is potentially a possibility for three-dimensional autonomous systems. 
 
 The remaining trajectories are those in the invariant plane; these are the flows of a two-dimensional autonomous subsystem that we now investigate.

\subsection{The invariant plane}\label{subsec:IP}

For $A>0$ the upper-half-ball phase space ($\tilde V\ge0$ and $z>0$) is divided into two disjoint parts by the invariant plane 
\be\label{inv.plane}
y= - \hat\lambda\sin\theta\, . 
\ee
Trajectories that pass through a point on one side of the plane must stay on that side, although they may start or end on a fixed point on it.  Moreover, $y$ is either monotonically increasing or monotonically decreasing on any such trajectory, depending on which side of the plane it lies. It follows from this that all fixed points with $z>0$ must lie on the invariant plane \eqref{inv.plane}, in which case they must also be fixed points of the 2-dimensional  autonomous subsystem found 
by using  \eqref{inv.plane} to eliminate the variable $y$ from \eqref{3Dsystem}:
\bea\label{2Dsystem}
x' &=& \hat\mu A- \hat\mu x^2 - B z^2 - xz^2 \nonumber \\
z' &=& z\left[ A + (\hat\lambda\cos\theta) x -z^2\right]\, ,  
\eea
where
\be\label{Bdef}
B= \hat\mu+\hat\lambda\cos\theta \equiv \hat\mu (1+X)\, . 
\ee
Recall that $A>0$ by assumption in this subsection,  but $B$ can have either sign; notice that 
\be 
B>0 \quad \Leftrightarrow \quad 1+X \ >0\, . 
\ee
The fixed points of the system \eqref{2Dsystem} with $z>0$  have coordinates given 
by the simultaneous solutions of the following two equations:
\be\label{2eqs}
(x + \hat\lambda\cos\theta)(Bx+A)=0\, , \qquad  z^2 = A+ (\hat\lambda \cos\theta) x \, . 
\ee
There are therefore two of these fixed points, generically, determined by the solutions of the quadratic equation for $x$:
\begin{itemize}
    
\item $x= - \hat\lambda\cos\theta$. For $\hat\lambda<1$ this gives us a fixed point of the full 3-dimensional system at 
\be
x+iy=- \hat\lambda e^{i\theta} \,,  \qquad z= \sqrt{1-\hat\lambda^2} \equiv \frac{\sqrt{-\Delta}}{2\alpha}\, , 
\ee
which implies $\tilde V=0$; this is a boundary ($\dot\chi=0$) fixed point, which exists (for $z>0$) only if $\Delta<0$ (which implies $A>0$ from \eqref{A>0}).

\item $Bx= - A$. Provided that $B>0$, this gives us a fixed point with
\be\label{2nd}
x= - \frac{A}{B} \, , \qquad y= - \hat\lambda \sin\theta\, , \qquad 
z= \sqrt{\frac{\hat\mu A}{B}} \, , 
\ee
which yields $\tilde V= (A/B^2)K$, where 
\be\label{Kdef}
K = \hat\lambda^2 + \hat\mu\hat\lambda\cos\theta -1  \equiv \frac{1}{(2\alpha)^2}\left[ \Delta + X \mu^2\right]\, . 
\ee
The condition $\tilde V\ge0$ is satisfied provided $K\ge0$, and we need strict inequality 
for the existence of an interior fixed point. Notice that 
\be
K>0 \quad \Leftrightarrow \quad \Delta + X\mu^2  \ >0\, . 
\ee
As $K\to0$  (for fixed $B>0$) on some curve in parameter space of decreasing $K$, the interior fixed point migrates to the boundary, where it coincides with the boundary fixed point (thereby creating a single degenerate boundary fixed point); further on the curve in parameter space, where $K<0$,  the interior fixed point is outside the upper-half-ball, leaving us with the original (non-degenerate) boundary fixed point only.

Combining the conditions $A>0$ and $K>0$, one finds that $X(X+1)>0$; this implies $X>0$ since $B>0$ is equivalent to 
$X+1>0$. It follows that there is no interior fixed point when $X\leq 0$, {\it i.e.} when $\cos\theta\le0$, although 
there may or may not be a boundary fixed point, according to whether $\hat\lambda<1$ or $\hat\lambda\ge 1$, respectively. 

\end{itemize}

 \begin{figure}[h!]
 \centering
 \begin{tabular}{cc}
 \includegraphics[width=0.45\textwidth]{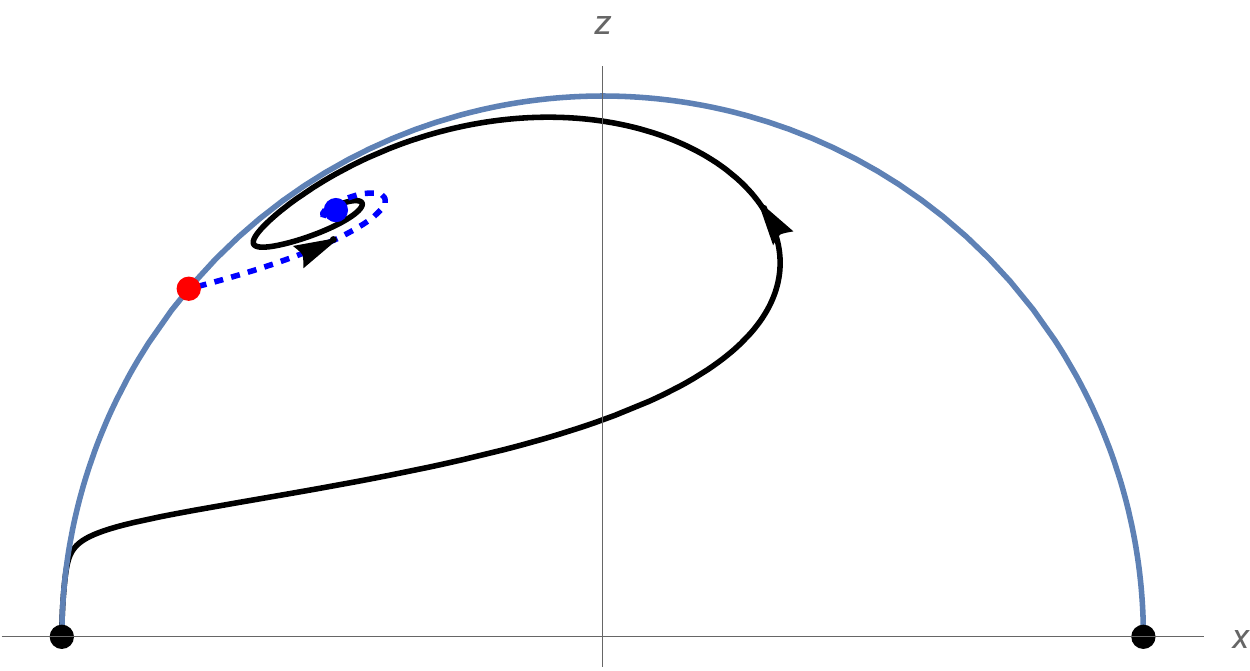}
 &
 \qquad \includegraphics[width=0.45\textwidth]{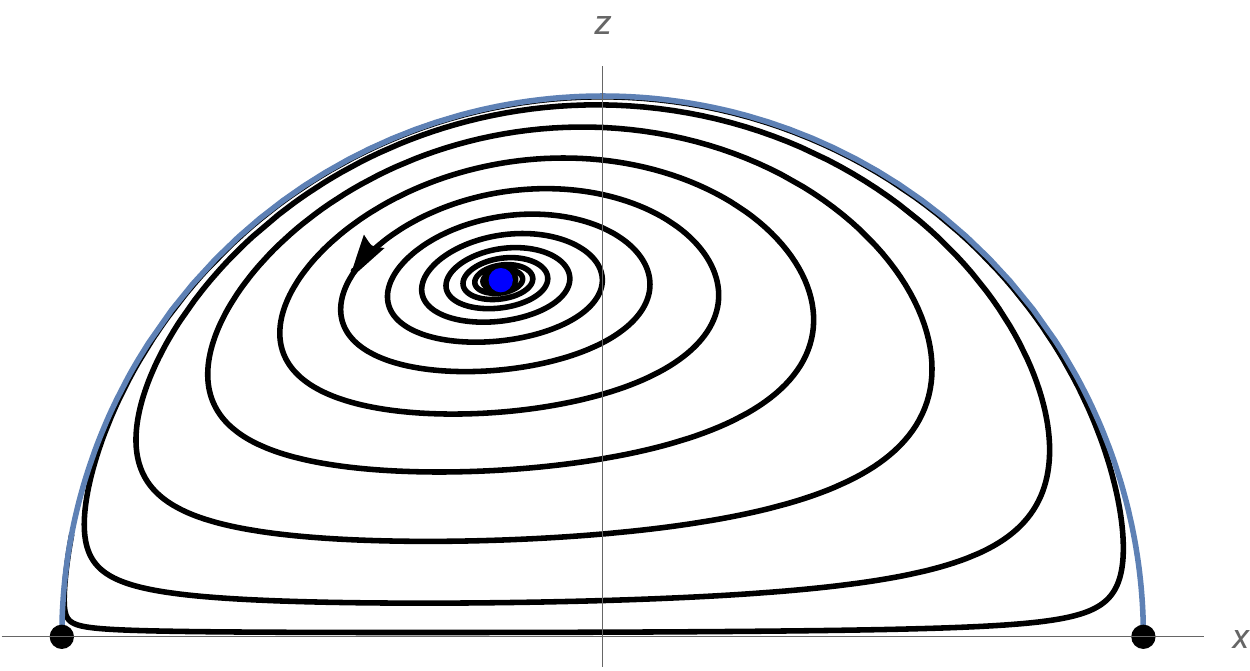}
 \\ (a)&(b)\\
 \includegraphics[width=0.42\textwidth]{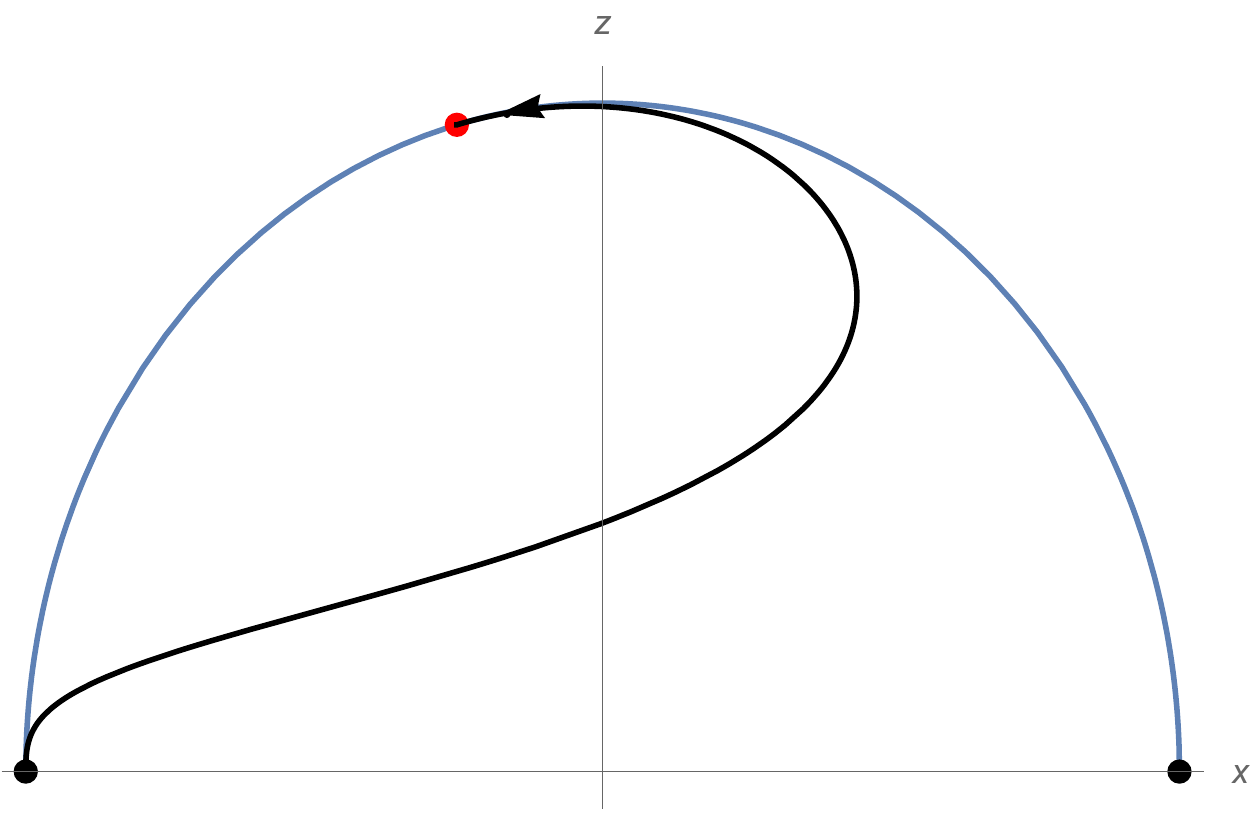}
 &
 \qquad \includegraphics[width=0.42\textwidth]{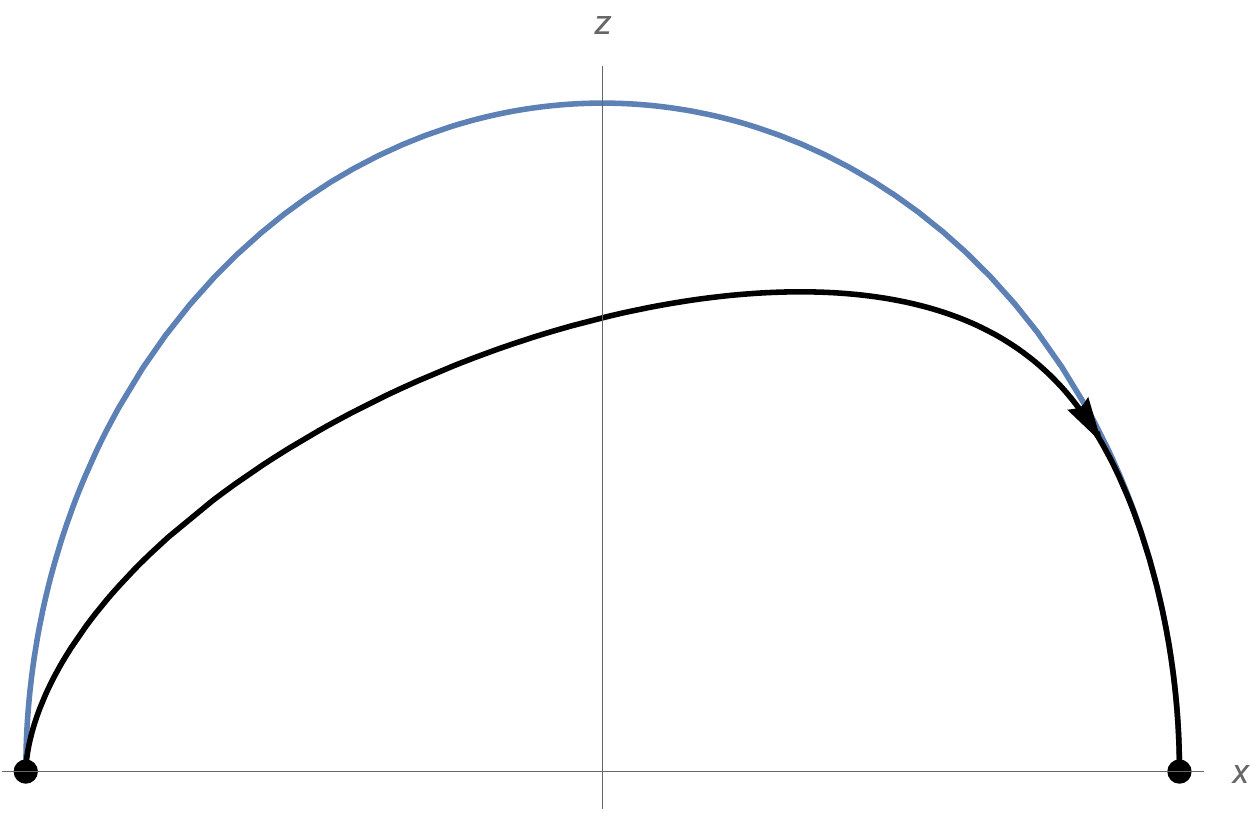}
 \\(c)&(d)
 \end{tabular}
 \caption
 {Trajectories lying in the invariant plane.
  (a) $\mu=2$, $\theta=\pi/6$, $\lambda=1.4$: the boundary fixed point is repulsive, the interior fixed point is attractive. The dotted trajectory is the unique separatrix from the
  boundary to the interior fixed point. (b) Same $\mu,\, \theta$, $\lambda=3$. There is only an attractive interior fixed point.(c) 
  Same $\mu,\, \theta$, $\lambda=1/2$. In this regime
 there is only a boundary point. (d) $\mu=2$, $\theta=3\pi/4$, $\lambda=2$: there is an invariant plane but there is not a boundary nor an interior fixed point.}
 \label{planacd}
 \end{figure}

As  the topology of  trajectories in the invariant plane will depend only on the existence of fixed points in the invariant plane
\eqref{inv.plane}, we see from the above analysis that there are four possible outcomes for trajectory topologies,
depending on the nature of the $z>0$ fixed points (assuming $A>0$ since otherwise the trajectories are 
not in the allowed region). Notably, the conditions for the existence of these fixed points can be expressed, in all four cases,  in terms of the $d$-independent variables $(\mu,\Delta, X)$, which implies $d$-independence of the topology of trajectories: 

\begin{itemize}

\item[(a)] A boundary fixed point and an  interior fixed point. This occurs when 
\be
\label{infixx}
\Delta <0  \qquad \& \qquad \Delta +X\mu^2  >0\, . 
\ee
[These two conditions imply $B>0$ (and $A>0$)]

\item[(b)] An interior fixed point only. This occurs (given $A>0$) when
\be
\Delta \ge 0 \qquad \& \qquad \Delta +X\mu^2 \, >0 \qquad\&\qquad  1+ X \, >0\, .
\ee

\item[(c)]  A boundary fixed point only. This occurs when 
\be
\Delta<0  \qquad \& \qquad \Delta + X\mu^2  \, \le0 \, . 
\ee

\item[(d)]  Neither a boundary fixed point nor an interior fixed point. This occurs (given $A>0$) when 
\be
\Delta \ge0 \qquad\&\qquad  X\le0 \ .
\ee

\end{itemize}
These four possibilities are illustrated in Figs. \ref{planacd}a,b,c,d.

Cases (a) and (b) are qualitatively the same as two cases discussed in \cite{Sonner:2006yn}; in case (b) the intersection of the invariant plane with the 
boundary of the upper-half unit ball and the $z=0$ plane forms a heteroclinic cycle of the autonomous system defined by \eqref{2Dsystem}, to which
all interior trajectories in the invariant plane are asymptotic as $\tilde\tau\to -\infty$ (which is equivalent to $t\to-\infty$). For $\sin\theta=0$ this means that if we start sufficiently far back in time the forward evolution in case (b) corresponds to a universe that passes arbitrarily close to $z=1$; as we discuss in the following section, this corresponds to a transient de Sitter-like phase. For $\sin\theta\ne0$, the interior trajectories on the invariant plane all have $z^2\le A<1$, so de Sitter-like phases can occur on these trajectories only if $\hat\lambda^2\sin^2\theta \ll 1$. 
There are trajectories {\it not in the invariant plane}, to be discussed in the following subsection, which do pass arbitrarily close to $z=1$, but not infinitely often.  

In cases (a) and (b) the late-time behaviour for trajectories in the invariant plane  is determined by the interior fixed point; in case (c) it is determined by the boundary fixed point and in case (d) all trajectories end on one of the two fixed points on the $z=0$ boundary of the invariant plane. As we shall see, these late-time possibilities apply more generally to trajectories of the full three-dimensional dynamical system defined by \eqref{3Dsystem}.

\subsection{Linearization at fixed points}

\begin{figure}[h!]
\centering
\includegraphics[width=0.6\textwidth]{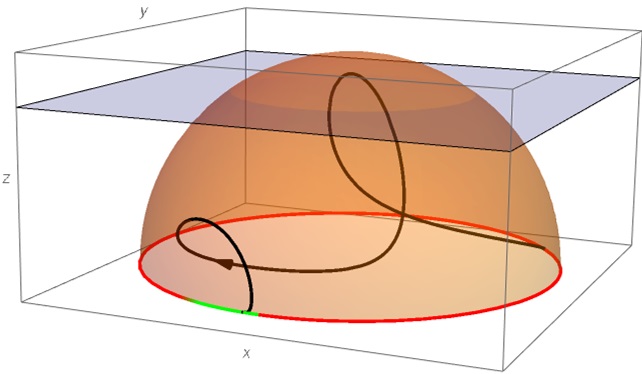}
\caption{Trajectory for $\mu=2$, $\lambda=4$, $\theta=\pi/4$, $d=4$. The trajectory
passes through the region $z>\sqrt{2/3}$ where the cosmology is accelerating, indicated by the horizontal blue plane, and then falls into the attractive (green) arc. For these values of the parameters, there are no other fixed points.}
\label{primera}
\end{figure}

\begin{figure}[h!]
 \centering
 \begin{tabular}{cc}
 \includegraphics[width=0.45\textwidth]{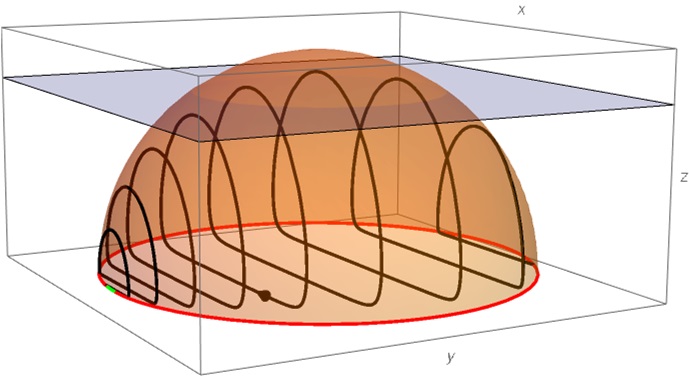}
 &
 \qquad \includegraphics[width=0.45\textwidth]{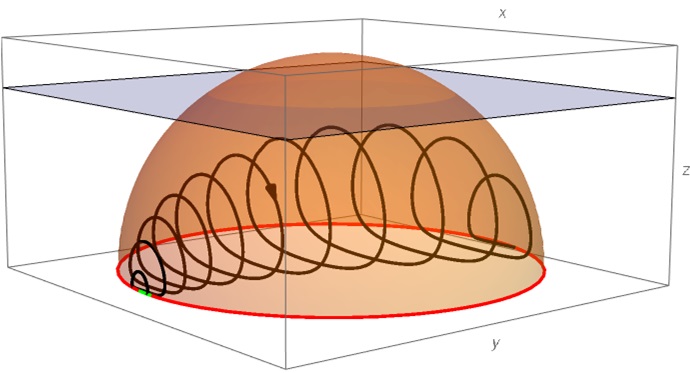}
 \\ (a)&(b)
 \end{tabular}
 \caption
 {Trajectories for $\mu=2$, $\lambda=20$, $\theta=\pi/16$. (a) The trajectory
goes through the region $z>\sqrt{2/3}$ a number of times,
then spirals into the attractive arc. (b) For a different initial condition, the trajectory stays in the region  $z<\sqrt{2/3}$ representing a decelerating cosmology.}
 \label{segunda}
 \end{figure}

To study the stability properties around a fixed point with $(x,y,z)= (x_0,y_0,z_0)$ we define
the new `shifted' variables
\be
\xi_1=x -x_0 ,\qquad \xi_2=y-y_0   ,\qquad \xi_3= z -z_0\, . 
\ee
Linearization about these fixed points yields equations of the form 
\be
\xi'_i =  L_{ij} \xi_j + \mathcal{O}(\xi^2)\, .
\ee
We shall first consider the boundary and interior fixed points, and then the circle of fixed points 
on the $z=0$ plane: 
\begin{itemize}
    \item Boundary fixed point. In this case 
\be
L = \left(\begin{array}{ccc} 
\hat\lambda^2 + 2\hat\mu\hat\lambda\cos\theta -1
& 2\hat\mu \hat\lambda \sin\theta & - 2\hat\mu\sqrt{1-\hat\lambda^2} \\
0 & \hat\lambda^2-1 & 0 \\ 
(\sqrt{1-\hat\lambda^2})  \hat\lambda\cos\theta  
& (\sqrt{1-\hat\lambda^2}) \hat\lambda\sin\theta & 2(\hat\lambda^2-1) \end{array} \right) \, . 
\ee
The eigenvalues are 
\be\label{evals1}
\hat\lambda^2-1\, , \qquad \hat\lambda^2-1\, , \qquad  2K\, ,  
\ee
where $K$ is defined in \eqref{Kdef}. The fact that there are two equal eigenvalues is a consequence of an SO(2) symmetry of the boundary dynamical system \eqref{bndry} when linearized about the boundary fixed point. 
The fixed point is a sink ({\it i.e.} an attractor) for $K<0$ but a source for $K>0$. 
The sign of the third eigenvalue depends on whether there is an interior fixed point. By comparison with 
\eqref{infixx}, we see that the third eigenvalue is positive whenever there exists an interior fixed point; in this case there
is a unique (separatrix) trajectory from the boundary fixed point to the interior fixed point. 

\item Interior fixed point. In this case
\be
L = \left(\begin{array}{ccc} 
\hat\mu A/B & 2\hat\mu\hat\lambda\sin\theta & 
-2\sqrt{\hat\mu A/B}\left(\hat\mu + K/B\right) \\
0 & - \hat\mu A/B & 0 \\
(\sqrt{\hat\mu A/B}) \hat\lambda\cos\theta & 
(\sqrt{\hat\mu A/B}) \hat\lambda\sin\theta & -2\hat\mu A/B
\end{array}\right)\, ,  
\ee
where (we recall) 
\be 
K = \hat\lambda^2 + \hat\mu\hat\lambda \cos\theta -1\, .  
\ee
As we have already seen, $K>0$ whenever there is an interior fixed point. The eigenvalues are now
\be
-\frac{\hat\mu A}{B} \, , \quad -\frac{\hat\mu A}{2B} - \Xi\, , \quad -\frac{\hat\mu A}{2B} + \Xi\, ,  
\ee
where 
\be
\Xi=\sqrt{\left(\frac{\hat\mu A}{2B}\right)^2 - 
\frac{2\hat\mu A K}{B}} \, . 
\ee
Notice that when $K=0$, the eigenvalues coincide with those of \eqref{evals1}, as expected since the interior 
and boundary fixed points coincide when $K=0$; this also explains the zero eigenvalue for $K=0$ because the fusion of two non-degenerate fixed points produces a degenerate fixed point.  For $K>0$ all three eigenvalues are negative when $\Xi$ is real, implying that the interior fixed point is an attractive node.
When $\Xi$ is imaginary, this fixed point becomes an attractive focus/spiral (since the real part of all three eigenvalues is negative).

\item The circle of fixed points at $z=0$.  To linearize about a fixed point on this circle with $x+iy= e^{i\omega}$ and $z=0$, we set
    \be
    x= \cos\omega + \xi \, , \qquad y= \sin\omega + \zeta\, , 
    \ee
    and then expand the equations of \eqref{3Dsystem} to first order in $(\xi,\zeta, z)$. The eigenvalues of this linear system are found to be 
    \be\label{autoarco}
    \{0\, , \  -2\hat\mu\cos\omega, \ 1+ \hat\lambda\cos(\omega- \theta)\}\, .
    \ee
    The zero eigenvalue tells us that the fixed points are degenerate, as expected since they are not isolated. Those fixed points for which $\omega$ satisfies
    \be
    \cos\omega >0\ \quad \& \ \quad 1+ \hat\lambda\cos(\omega- \theta) < 0
    \ee
    are endpoints for all trajectories in the upper half ball.  They form an arc of the circle
    $x^2+y^2=1$, as can be seen from the numerics in figures \ref{primera} and \ref{segunda}.

\end{itemize}

\begin{figure}[h!]
\centering
 \begin{tabular}{cc}
 \includegraphics[width=0.45\textwidth]{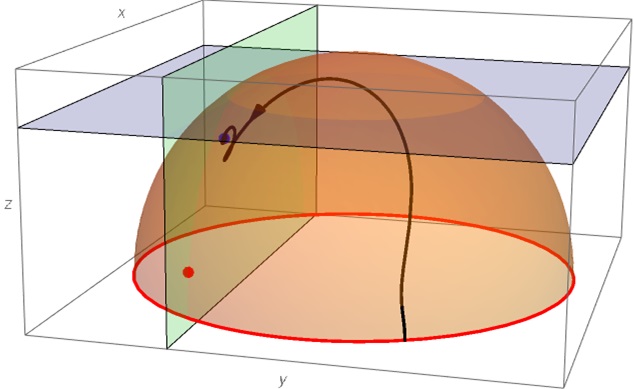}
 &
 \qquad \includegraphics[width=0.45\textwidth]{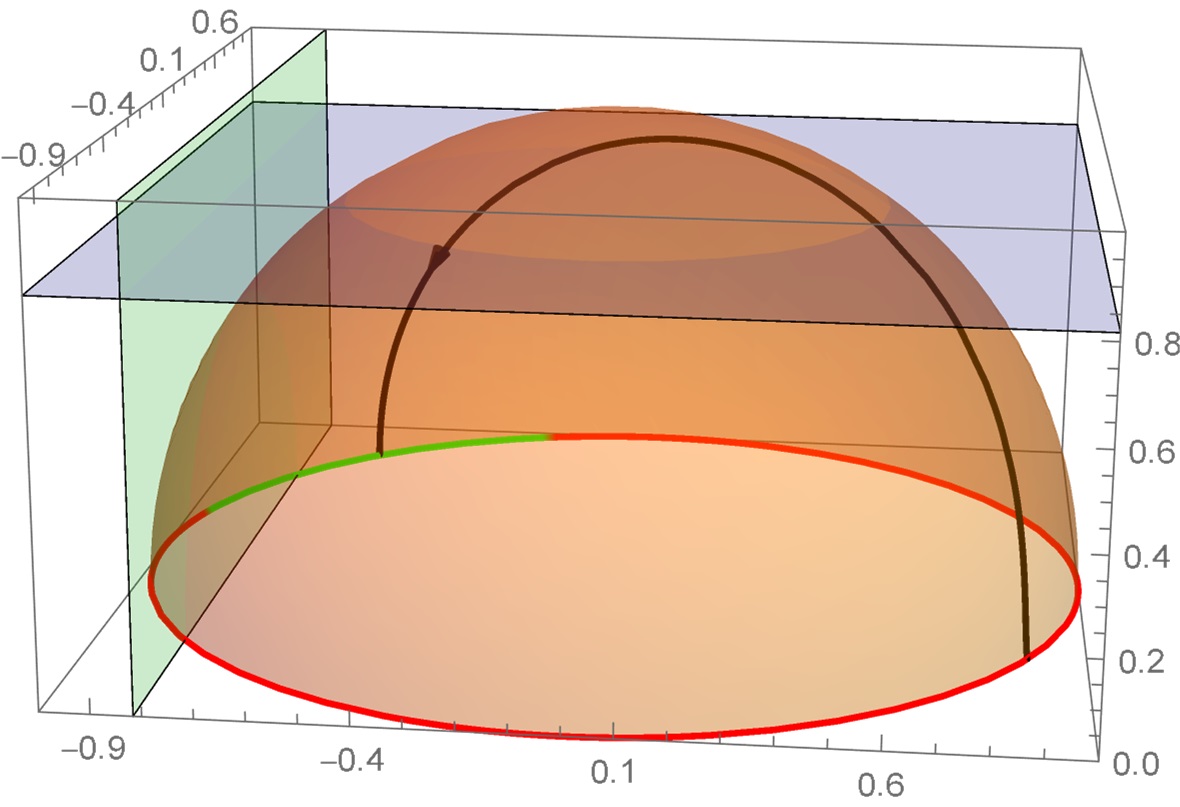}
 \\ (a)&(b)
 \end{tabular}
\caption{(a) Trajectory for $\mu=2$, $\lambda=1.7$, $\theta=\pi/6$. Both the interior (blue) and boundary (red)  fixed points lie on the invariant (vertical) plane. As the boundary fixed point has repulsive directions, the trajectory falls into the attractive, interior fixed point. 
(b) $\mu=2$, $\lambda=2$, $\theta=3\pi/4$. This is a regime where there is no boundary nor an interior fixed point, but there is still an invariant plane. Trajectories go to the attractive arc, without crossing the plane.
}
\label{tercera}
\end{figure}

\begin{figure}[h!]
 \centering
 \begin{tabular}{cc}
 \includegraphics[width=0.45\textwidth]{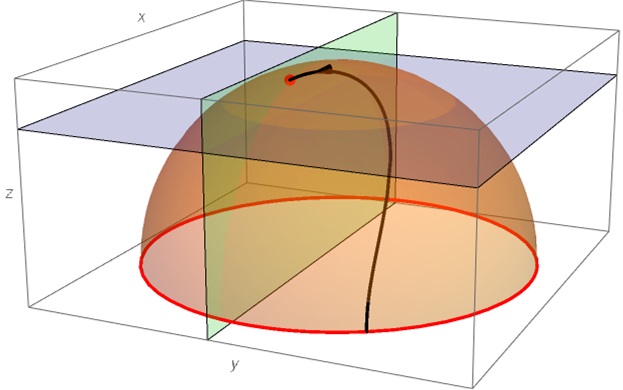}
 &
 \qquad \includegraphics[width=0.45\textwidth]{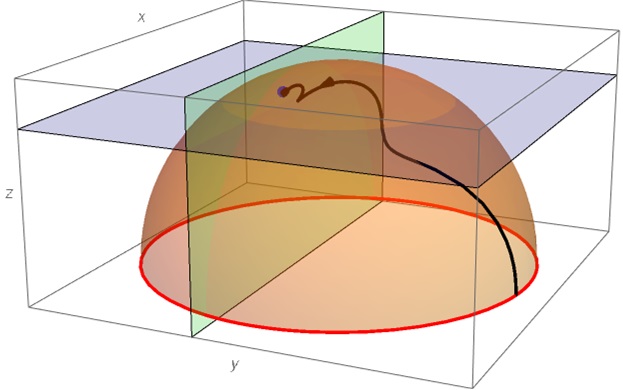}
 \\ (a)&(b)
 \end{tabular}
 \caption
 {(a) Trajectory for $\mu=2$, $\lambda=1/2$, $\theta=\pi/6$. In this regime, there is only one (attractive) boundary fixed point. Since $\lambda<\lambda_c=1$, it lies in the  region $z>\sqrt{2/3}$.
(b)  Trajectory for $\mu=8$, $\lambda=2$, $\theta=\pi/16$. There is only an interior attractive fixed point, lying in the region $z>\sqrt{2/3}$.
Both cosmologies (a) and (b) 
describe eternal acceleration.}
 \label{cuarta}
 \end{figure}

Using the above observations, our earlier analysis of the various possible phase-plane portraits for the autonomous dynamical system 
on the invariant plane of constant $y$ may be extended to portraits for the full 3-dimensional system.   Some examples are shown in Figs. \ref{primera}--\ref{cuarta}.

Some notable features are:

\begin{itemize}

\item When there is an invariant plane, and either a boundary fixed point, interior fixed point or both, then all almost all trajectories end on the interior fixed point, when this exists, or the boundary fixed point otherwise. All trajectories not on the invariant plane start on the circle of fixed points at $z=0$. 

\item When there is an invariant plane but neither an interior nor a boundary fixed point on it, trajectories start at one $z=0$ fixed point and end on another one on the same side of the invariant plane. The starting and end points are on complementary arcs of the 
unit circle on the $z=0$ plane which include the starting point and end point of all trajectories on the invariant plane. 

\item As the trajectories fill the possible phase-space, there are trajectories that approach arbitrarily close to $z=1$ (in addition to the one boundary trajectory that passes through this point). 
For sufficiently small $\theta$, and sufficiently large $\mu$ or $\lambda$, there are trajectories that pass arbitrarily close to $z=1$ many times. 

 \end{itemize}

\medskip

Finally, we comment on the special case $\theta =\frac{\pi}{2}$, {\it i.e.} $X=0$, for which the 
potential term preserves the full $Sl(2;\mathbb{R})$ invariance of the massless theory. 
As the existence of an interior point requires $\cos\theta>0$, this is also the limiting 
case for which the interior point ceases to exist. The topology of trajectories can be deduced 
from following observations:

\begin{itemize}
    
    \item For $\hat \lambda\leq 1$, there is an attractive boundary fixed point lying on the invariant plane. The linearized system has eigenvalues $(1-\hat\lambda^2)\{ -1,-1,-2\} $.
    This is a sink for all trajectories.
    
    \item For $\hat \lambda>1$, all trajectories start and end on points of the circle $x^2+y^2=1,\ z=0$.

\end{itemize}
Although the scalar field $\phi_1$ that couples to the axion does not appear in the scalar potential when $\theta =\frac{\pi}{2}$, the fields $(\chi,\phi_1)$  remain gravitationally coupled to $\phi_2$, through the Friedmann constraint in the context of FLRW cosmology. This makes it difficult to find explicit analytic solutions, although these can be found for {\it boundary} trajectories when $\dot \phi_1=0$ since in this case the model reduces to a gravity-dilaton system with an exponential potential, for which the  general solution is known \cite{Russo:2004ym}.

 \section{Equation of State and Cosmic Acceleration}\label{sec:eos}

We have been investigating cosmological solutions for a particular example of Einstein gravity in $d$ spacetime 
dimensions coupled to scalar fields. The Lagrangian density for the generic model of this type, for an appropriate 
choice of units, takes the form 
\be\label{grav-scalars}
\mathcal{L} = \frac12\sqrt{-\det g}\left\{ 2R -  g^{\mu\nu}\partial_\mu\Phi^\alpha \partial_\nu\Phi^\beta h_{\alpha\beta}(\Phi) - V(\Phi)\right\} \, , 
\ee
where $h$ is a Riemannian metric for a target space with local coordinates  $\{\Phi^\alpha; \alpha = 1\dots,n\}$ and $V$ is function 
on this space; the scalar fields are maps from spacetime to the target space.  In our case, the target space is the homogeneous space $SL(2; \mathbb{R}) \times \mathbb{R}$, with coordinates $\{\vec\phi,\chi\}$  and $V$ is an exponential function of a linear combination of the two components of $\vec\phi$ only. 

The Einstein field equations, 
\be\label{Einstein}
R_{\mu\nu}(g) =T_{\mu\nu} -\frac{1}{(d-2)}\, g_{\mu\nu} \, g^{\rho\sigma}T_{\rho\sigma}\, , 
\ee
determine the spacetime Ricci tensor in terms of the scalar field stress-energy tensor
\be
T_{\mu\nu} = \frac12\left\{ \partial_\mu\Phi^\alpha \partial_\nu\Phi^\beta h_{\alpha\beta} - \frac12 g_{\mu\nu}\left[g^{\rho\sigma} \partial_\rho \Phi^\alpha \partial_\sigma\Phi^\beta h_{\alpha\beta} + V\right] \right\} \, . 
\ee
For an FLRW spacetime with metric 
\be
ds^2 = g_{00}(\tau)\,  d\tau^2  + g_{ij}(\tau)\,  dx^i dx^j \, , 
\ee
and scalar fields $\Phi$ that are functions only of the time coordinate $\tau$, the only non-zero components of 
the stress-energy tensor are
\be
T_{00} = -g_{00}\, \rho \, , \qquad T_{ij} = g_{ij}\, p \, , 
\ee
where $\rho$ and $p$ are, respectively, the energy density and pressure of the scalar-field matter:
\be
\rho= -\frac{1}{4g_{00}} \left[ |\dot\Phi|^2 - g_{00}V\right]\, , \qquad p= -\frac{1}{4g_{00}} \left[ |\dot\Phi|^2 + g_{00}V\right]
\ee
with 
\be
|\dot\Phi|^2 := \frac{d\Phi^\alpha}{d\tau} \frac{d\Phi^\beta}{d\tau}  h_{\alpha\beta} \, . 
\ee
For the same FLRW metric assumption we now have
\be\label{rzz}
R_{00} =\frac12\left\{ |\dot\Phi|^2 + \frac{1}{(d-2)} g_{00} V\right\}\, . 
\ee

If we write the FLRW metric in the standard form
\be
ds^2= -dt^2 + S^2(t) \bar g_{ij}dx^i dx^j \, ,  
\ee
for time-independent maximally-symmetric $(d-1)$-metric $\bar g$, then a direct calculation yields 
\be
R_{00} = -(d-1) \frac{\partial^2_t S}{S} \, , 
\ee
which shows that the expansion of the FLRW universe is accelerating iff $R_{00} <0$.
Let us note here that the scale factor $S(t)$ is related to the variable $\varphi(\tilde\tau)$ used
for the global phase-space analysis of the previous section (and $t$ is related to $\tilde\tau$) by 
\be
S(t) = e^{\beta\varphi(\tilde\tau)}\, , \qquad 
2\alpha\,  dt= \left(\frac{z}{m}\right) e^{- \frac12 \vec\lambda \cdot\vec\phi} d\tilde\tau\, .  
\ee

\subsection{The universal one-axion model}

For our one-axion model with Lagrangian density \eqref{basicmodel} and  the specific FLRW metric of \eqref{FLRW} with our choice of 
time parametrization determined by \eqref{tiempotau}, we have
\be
g_{00} = - e^{- \vec\lambda\cdot\vec\phi}\, ,  \qquad V= m^2 e^{\vec\lambda\cdot\vec\phi} \, , 
\ee
and 
\be
|\dot\Phi|^2 = |\vec u|^2 + e^{\vec\mu\cdot\vec\phi} \dot\chi^2 = v^2-m^2 \, ,  
\ee
where $\vec u = \partial_\tau \vec\phi$, and the second equality uses the constraint \eqref{constraint} (with $v= \dot\varphi$). 
We thus find that
\be
\rho= \frac14 v^2 e^{\vec\lambda\cdot \vec\phi} \, , \qquad 
p=  \frac14  \left(v^2-2m^2\right)  e^{\vec\lambda\cdot \vec\phi} \, , 
\ee
and hence the equation of state
\be\label{eos}
w:= \frac{p}{\rho} = 1- 2z^2 \qquad (z= m/v) \, . 
\ee
Notice that $z=1$ gives $w=-1$, which is the equation of state corresponding to a cosmological constant\footnote{Equivalently, it is the equation of state of an ultra-relativistic tensile material, which could be interpreted as a space-filling brane \cite{Townsend:2021wrs}.} while $z=0$ gives $w=1$, which is the ``stiff matter'' equation of state.  

From \eqref{rzz} we find that 
\be
R_{00} = \frac{m^2}{2z^2} \left[ 1 - \left(\frac{d-1}{d-2}\right) z^2\right] \, , 
\ee
and hence that 
\be\label{accel}
R_{00} <0 \quad \Leftrightarrow \quad z > \sqrt{\frac{d-2}{d-1}} \quad 
\Leftrightarrow\quad w < - \left(\frac{d-3}{d-1}\right)\, ,  
\ee
which is the condition for accelerated expansion. This includes de Sitter expansion ($w=-1$) but we are also interested  here in the de Sitter-like expansion that one gets when $w$ approaches $-1$ during a period of transient acceleration; {\it i.e.} a de Sitter-like phase.  From \eqref{eos} we see that 
\be
z= 1-\frac14 \delta \quad \Rightarrow \quad w= -1 + \delta + \mathcal{O}(\delta^2) \, , 
\ee
which provides a measure of how close a trajectory must get to the ``north pole'' of the unit upper-half ball in the global phase space ({\it i.e.} $z=1$) for the equation of state to approximate $w=-1$ to within some given error $\delta$.

\subsection{Late-time accelerated expansion}

Late-time cosmologies are determined by the behavior at fixed points of the three-dimensional dynamical system \eqref{3Dsystem}. From the formula \eqref{eos} we may determine the equation of state, {\it i.e.} the value of $w$, at the fixed  points. We then use \eqref{accel} to find the range of parameters for which the FLRW cosmology at the fixed point has accelerated cosmic expansion. We shall present these results both for the parameters $(\lambda,\mu,\theta)$ and the $d$-independent parameters $(\Delta,\mu,X)$. In translating from one set of parameters to the other it is important to appreciate that the identity $\cos^2\theta\le1$ becomes the inequality 
\be\label{phys-R}
\Delta\ge X^2\mu^2 - 4\alpha^2\, , 
\ee
which therefore defines the physical region of the $(\Delta,\mu,X)$ parameter-space. 
We now consider the fixed points in turn:
\begin{itemize}

\item Fixed points on $x^2+y^2=1$ circle ``at infinity''. In this case $z=0$ and hence
\be
w=1\, .   
\ee
This yields the ``stiff matter'' equation of state, corresponding to negligible scalar potential energy.

\item Boundary fixed point; this exists for $\hat\lambda<1$ (equivalently, $\Delta<0$) and is a late-time attractor in the absence of an interior fixed point. At this fixed point $z^2= 1-\hat\lambda^2$, which implies
\be
w= -1 + 2\hat\lambda^2 \, ,  
\ee
and hence accelerated expansion of the corresponding FLRW cosmology when 
\be 
\label{poww}
\hat\lambda <\frac{1}{\sqrt{d-1}} \qquad \left(\Leftrightarrow \quad \lambda < \sqrt{\frac{2}{d-2}} \equiv \lambda_c\right) \, .  
\ee
Notice that this condition for acceleration at the fixed point implies the condition for the existence of the fixed point, so the condition
for acceleration at a boundary fixed point coincides with the standard result for a dilaton with exponential potential ($\lambda$ must be less than the critical value $\lambda_c$). This was to be expected  since the boundary of the phase-space of the dynamical system corresponds to constant axion field. Notice that 
\be
\lambda <\lambda_c \quad  \Leftrightarrow  \quad\Delta<-2\ .
\ee

\item Interior fixed point. This exists provided $A>0$, $B>0$ and $K>0$, and these three conditions are jointly equivalent to 
\be
\label{gyyy}
\Delta >-X\mu^2\, , \qquad X\ge 0 \,, \qquad \Delta< X^2\mu^2\, .   
\ee
At this fixed point we have $z= \sqrt{\hat\mu A/B}$, and hence
\be
w= 1 - \frac{2\hat\mu A}{B}\, ,  
\ee
so that accelerated expansion occurs for 
\be
\label{bbpon}
\frac{\hat\mu A}{B} > \frac{d-2}{d-1} \, .  
\ee
Using the expressions for $A$ and $B$ in \eqref{A>0} and \eqref{Bdef}, we find that this is equivalent to 
\be\label{Quad-lambda}
\lambda^2\sin^2\theta + \frac{2\lambda}{\mu} \cos\theta  <  \lambda_c^2 \, ,  
\ee
which shows that accelerated cosmic expansion is possible for arbitrarily large $\lambda$ if $\theta$ is sufficiently small and $\mu$ sufficiently large. In terms of the $d$-independent parameters, the equivalent inequality is 
\be
\label{intebb}
\Delta <  X^2\mu^2-2X-2\ .
\ee
The boundary of the region defined by this inequality can be viewed as a $\mu$-dependent parabola in the $X-\Delta$ plane, which we shall call $P$.
The region $R$ in this plane for which the interior fixed-point cosmology is accelerating lies under $P$ but we must also take into account (i) the conditions \eqref{gyyy} for the interior fixed point to exist, and (ii) the condition \eqref{phys-R} for physical $(\Delta,\mu,X)$ parameters. 
We consider these in turn:
    
\begin{enumerate}

   \item  As the inequality $\Delta<X^2\mu^2$ (or $A>0$) is implied by $X\ge0$ (or $\cos\theta\ge0$), given \eqref{intebb}, 
  the only fixed-point-existence conditions on $R$ that we need to consider are $X\ge0$ and $\Delta>-X\mu^2$ (or $K>0$) and these imply, given \eqref{intebb}, that
     \be
    X\mu^2> 2 \qquad  \left(\Leftrightarrow \quad \vec\mu\cdot\vec\lambda >2\right) \, , 
    \ee
    which is stronger than $X>0$ and hence replaces it.  As the parabola $P$ and lines $\Delta= - X\mu^2$ and $X=2/\mu^2$ have a common intersection at $\Delta=-2$, the point with coordinates
    \be
    X= \frac{2}{\mu^2}, \qquad \Delta=-2
    \ee
    is a boundary point of $R$, and all points with $X<2/\mu^2$ lie outside $R$. 
    
     \item  The `acceleration parabola' $P$  intersects  the parabolic boundary of the physical region of parameters defined 
     by \eqref{phys-R} at the point
  \be
    X= 2\alpha\beta =\frac{1}{d-2}\, , \qquad \Delta = 4\alpha^2 \left(\beta^2\mu^2 -1\right)\, .
   \ee
  This point is therefore on the boundary of $R$ and all points with $X>2\alpha\beta$ lie outside $R$. 
  
  \end{enumerate}
  Putting together the above conditions for $X$ we see that the region $R$ lies in the strip of the $X-\Delta$ plane defined by 
   the inequalities
\be\label{muc}
2<X\mu^2 < 2(\mu/\mu_c)^2\, , 
    \qquad \mu_c = \sqrt{2(d-2)} \equiv \frac{2}{\lambda_c} \, . 
    \ee
 As this strip exists only when $\mu>\mu_c$ there can be no acceleration at an interior fixed point unless $\mu>\mu_c$.
This may be viewed as a strong-coupling requirement for acceleration at the interior fixed point. 

Taking all the above observations into account, we see that $R$ exists only for $\mu>\mu_c$, in which case it is the region 
defined by the inequalities \eqref{phys-R} and \eqref{intebb}, and the 
two further inequalities 
    \be
      \Delta>-X\mu^2 \, ,  \qquad X\mu^2 >2\, .  
     \ee
   In terms of the $(\lambda,\mu,\theta)$ parameters, this is the region satisfying the acceleration inequality \eqref{Quad-lambda}
   subject to the two further inequalities
   \be
   \lambda^2 + \mu\lambda\cos\theta -4\alpha^2 >0\, , \qquad \mu\lambda \cos\theta >2\, . 
   \ee
\end{itemize}

For the purposes of the following subsection, we now ask and answer the following question: what is the region in parameter space for which there is an eternally accelerating cosmology either at a boundary fixed point or at an interior fixed point, or both? We shall answer this question using the $(\Delta, \mu,X)$ parameters, and then present the equivalent answer in 
$(\lambda,\mu,\theta)$ parameters. The answer for $\mu\le\mu_c$ is obvious because an accelerating cosmology is only possible at a boundary fixed point, so the region we seek  in this case is the physical region of the $X-\Delta$ plane for which 
\be
\Delta <-2   \qquad (\mu\le \mu_c)\, . 
\ee

For $\mu>\mu_c$ the region still includes $\Delta<-2$ but it now also includes those points 
in the $X-\Delta$ plane with $\Delta\ge -2$ that lie under the segment of the parabola $P$ in the strip
\be 
\frac{2}{\mu^2} < X < \frac{2}{\mu_c^2}\, . 
\ee
In other words, 
\be\label{anti-swamp1}
\Delta<\left\{\begin{array}{ccc} -2 &&   (X\mu^2 \le 2) \\ X^2\mu^2 -2X-2 && (X\mu^2\ge 2) \, , \end{array} \right.
\ee
where it should be understood that we restrict to the physical region defined by \eqref{phys-R}. This proviso is unnecessary 
after expressing this inequality  in terms of the $(\lambda,\mu,\theta)$ variables, as we shall now do.

Recall that the acceleration inequality \eqref{intebb} is equivalent to \eqref{Quad-lambda}, which we may write as 
\be
Q(\lambda)<0 \, , \qquad Q(\lambda) := \lambda^2\sin^2\theta + \frac{2\lambda}{\mu} \cos\theta - \lambda_c^2\, . 
\ee
This inequality (which is needed only for $\cos\theta>0$) states that $\lambda\in (\lambda_- ,\lambda_+)$, where $\lambda_\pm$ are the two roots of $Q(\lambda)$. As $\lambda_-<0$, the inequality $Q<0$ is equivalent to  $\lambda<\lambda_+$ where
\be\label{lplus}
\lambda_+(\mu,\theta)  = \frac{1}{\mu \tan\theta\sin\theta} 
\left[ \sqrt{1+ \lambda_c^2 \mu^2\tan^2\theta} -1\right] \, .  
\ee
Using this result, we may re-express the inequality \eqref{anti-swamp1} for $\Delta$ as the following inequality for $\lambda$: 
\be\label{anti-swamp2}
\lambda < \left\{\begin{array}{ccc} \lambda_c && (\mu\lambda\cos\theta\le2) \\
\lambda_+(\mu,\theta) &&  (\mu\lambda\cos\theta\ge2)  \, .  \end{array} \right.
\ee

\subsection{Eternal acceleration and the swamp}

Let us return to the general gravity-coupled scalar-field theory defined by 
the Lagrangian density of \eqref{grav-scalars}. What constraints on this 
model are imposed by the requirement that it be derivable from string/M-theory 
as an effective theory? On the basis of an examination of a variety of examples, 
it was conjectured in \cite{Obied:2018sgi} that when the potential $V$ is positive it 
must satisfy 
\be\label{vafa}
|\boldsymbol{\nabla} V|/V > c = \mathcal{O}(1)\, . 
\ee
Violation of this bound by any particular gravity-scalar model would then relegate it to 
the ``swampland’’. As it stands, this swampland bound is imprecise but one can suppose that there is some underlying property of string/M-theory, or quantum gravity more generally, that leads to a bound of the above type but with a definite value for the constant $c$. 

Our proposal for this property is that the spacetime metric can never have a future cosmological event horizon; in the context of FLRW cosmology, this becomes 
a no-eternal-acceleration principle because any expanding universe with positive 
asymptotic accelerated expansion (de Sitter being one example) has a future cosmological event horizon. 
One motivation for this proposal is that there are various known difficulties in formulating quantum gravity in spacetimes with a future cosmological event horizon \cite{Witten:2001kn,Dyson:2002pf,Banks:2002wr}.  Another motivation is that the no-eternal-acceleration principle for FLRW cosmology implies a relation of the type \eqref{vafa}. For example, for a simple model with a single scalar field $\phi$ with a potential $V=m^2e^{\lambda\phi}$, it implies the bound $\lambda\ge \lambda_c$. Equality is allowed since zero acceleration in the $t\to\infty$ limit implies the absence of a future cosmological event horizon \cite{Boya:2002mv}.

To apply this no-eternal-acceleration principle to the universal one-axion model, we recall that 
 the region in which acceleration occurs either at a boundary fixed point or at an interior fixed point (or both) 
 can be defined as that $\mu$-dependent subregion of the physical region of the $X-\Delta$ plane that lies 
 below a particular continuous curve, which divides the physical region of this plane into two disjoint parts. To avoid the possibility
 of eternal acceleration we must choose parameters corresponding to  points of the $X-\Delta$ plane that lie on or above this 
 continuous curve. In the case that $\mu\le\mu_c$, this prescription yields the swampland bound 
 \be
 \Delta\ge -2 \qquad  (\Leftrightarrow \quad \lambda \ge \lambda_c) 
 \ee
This is expected because the axion field is constant at the boundary fixed point, which is the only one for which 
acceleration is possible when $\mu<\mu_c$. 
 
In the case that $\mu>\mu_c$, the same prescription yields the swampland bound 
\be
\Delta\ge \left\{\begin{array}{ccc} -2 &&   (X\mu^2 \le 2) \\ X^2\mu^2 -2X-2 && (X\mu^2\ge 2) \, , \end{array} \right.
\ee
where it should be understood that we restrict to the physical region of $X-\Delta$ plane for which \eqref{phys-R} is satisfied. This proviso
is unnecessary when the swampland conditions are expressed in terms of the $(\lambda,\mu,\theta)$ parameters, in which case
we have, for $\mu>\mu_c$, 
\be\label{swamp2}
\lambda \ge  \left\{\begin{array}{ccc} \lambda_c && (\mu\lambda\cos\theta\le2) \\
\lambda_+(\mu,\theta) &&  (\mu\lambda\cos\theta\ge2)  \, , \end{array} \right.
\ee
where the function $\lambda_+$ is given in \eqref{lplus}.
Let us examine the implications of the inequality $\lambda\ge \lambda_+$. It is equivalent to
\be\label{noacc}
\lambda^2\sin^2\theta + \frac{2\lambda}{\mu} \cos\theta \ge \lambda_c^2  \qquad (\cos\theta>0) \, . 
\ee
For fixed $(\lambda,\mu)$, this inequality takes the form $f(\theta) \ge \lambda_c^2$ where $f$ takes a minimum value of $2\lambda/\mu$  at 
$\theta=0$. It follows, given the other swampland condition $\lambda>\lambda_c$, that \eqref{noacc} is  trivially satisfied when 
$\mu\leq \mu_c$, as should be the case. We may therefore rewrite \eqref{noacc} as 
\be
\mu_c< \mu \leq \frac{2\lambda \cos\theta}{\lambda_c^2 -\lambda^2\sin^2\theta}\, .  
\ee
The second inequality is a swampland upper bound on $\mu$ in terms of $(\lambda,\theta)$, and as $\mu^2$ is a measure of the curvature of the hyperbolic scalar-field target space, we see that the \eqref{noacc} can be interpreted as a swampland constraint on this curvature.

Finally, we remark that if there were a constant $C$ such that $\lambda_+>C$ for all  $\mu$ and $\theta$ then it might be possible for \eqref{swamp2} to be consistent with a single swampland bound of the form 
$\lambda > c$, with $c$ the constant of \eqref{vafa}. In fact, the function $\lambda_+(\mu,\theta)$ has no upper 
bound but it does have an upper bound if $\theta>\theta_0$ for some fixed angle $\theta_0$, and this upper bound
is of order 1 unless $\sin\theta_0<<1$; under these circumstances, a single bound of the form $\lambda> c=\mathcal{O}(1)$ would suffice. However, not only would the precision of the bound \eqref{swamp2} be lost but so too would the derivation of swampland bounds from a principle of no eternal acceleration, since this principle allows arbitrarily small $\theta$.  On the other hand, we do find in the following section that $\theta$ is greatly restricted by maximal supersymmetry, and it may be that a lower bound on $\theta$ is a general feature of  
string/M-theory compactifications.



\section{Application to maximal massive supergravity}
\label{applisec}

In this section we shall see how our universal one-axion model is a consistent truncation of 
maximal supergravity theories directly related to those that are effective theories for string/M-theory. 
We say `related' because we require  some toroidal compactification to $d=4$ that yields a ``massive'' 
deformation of  the (ungauged) N=8 supergravity.  The Cremmer-Julia
compactification of 11D supergravity, for example, yields the `massless' N=8 supergravity \cite{Cremmer:1979up}
but a modification to allow for a non-zero value of the 4D reduction of the 11D 4-form field-strength leads to 
a massive  deformation of N=8 supergravity with a positive exponential potential \cite{Aurilia:1980xj}; this was recently 
reviewed in \cite{Townsend:2019uzj}, where it was called the ANT construction.

Another massive deformation of N=8 4D supergravity, again with a positive exponential potential,  can be found by toroidal 
reduction of the 10D `mIIA' supergravity constructed by Romans \cite{Romans:1985tz}. The same massive 
N=8 4D supergravity can be found from a particular supersymmetry-preserving Scherk-Schwarz (SS) reduction\footnote{SS reduction exploits the existence of a $U(1)$ symmetry, but this breaks the 
local supersymmetry unless the fermion fields are $U(1)$-inert, which was not the case in the original application of the idea 
\cite{Scherk:1979zr,Cremmer:1979uq}.} of the 10D IIB supergravity; this is because there is already a coincidence in $d=9$ between the $S^1$ reduction  of mIIA and the SS compactification of IIB using the $U(1)$ symmetry that shifts the IIB axion field by a constant \cite{Bergshoeff:1996ui}, and a further $T^5$ reduction of the IIB theory therefore yields the same result 
as the $T^6$ reduction of mIIA.

Many more distinct massive deformations of N=8 4D supergravity can be found by similar SS reductions of IID 
supergravity  \cite{Cowdall:1996tw}: first we Kaluza-Klein (KK) reduce on $T^n$  to the maximal `massless' supergravity in $d=11-n$, 
with $n\ge2$ to ensure that there is an axion field for the next step, which is an SS reduction to a $(10-n)$-dimensional
massive maximal supergravity with an exponential potential.  Further KK reduction then yields a new
massive deformation of N=8 4D supergravity. There are 71 distinct massive maximal 4D supergravity theories
that can be found this way. One of them, equivalent to that found directly in 4D by the ANT construction,  comes from an 
SS reduction of the 5D maximal supergravity using the axion field obtained by dualization of the 4-form field-strength coming 
directly from 11D. However, the massive N=8 4D supergravity with the mIIA/IIB origin is not included in the count of 71, so 
the total number of distinct massive 4D N=8 supergravity theories with a simple exponential potential is 72 \cite{Cowdall:1996tw}. 

Many more massive supergravity theories can be found by `simultaneous' SS reduction using multiple 
axion-shift symmetries \cite{Cowdall:1996tw} or more complicated SS reduction 
ans\"atze \cite{Lavrinenko:1996mp,Kaloper:1998kr}, 
but then we have more complicated potentials depending on multiple independent mass parameters. The simplest 
way to find our universal one-axion model as a consistent truncation is to first set all but one of these mass parameters 
to zero, in which case we recover one of the 72 massive N=8 supergravity theories with an exponential potential. This
implies that no new universal one-axion model ({\it i.e.} with new parameters) can be found by consistent truncation of 
the multi-mass-parameter massive N=8 supergravity theories, so no generality is lost (for the purposes of this paper) 
by ignoring them. 

We shall see that consistent truncation of any massive N=8 4D supergravity to our universal one-axion model 
restricts the three dimension-dependent and dilaton-basis independent parameters of this model to one of 
two sets of values.  To arrive at this  conclusion we will need to investigate some special cases of massive maximal 
supergravity theories. We first consider the 4D massive N=8 supergravity of \cite{Aurilia:1980xj}; this will also be 
an opportunity to review aspects of the  work of  L\"u and Pope on maximal `massless' supergravity theories \cite{Lu:1995yn}, which already tells us that $\mu=2$, necessarily. 

Next, we investigate consistent truncations to the universal one-axion model of the $d=8,7$ massive maximal supergravity 
theories found in \cite{Cowdall:1996tw} by the method of SS reduction. In all these cases  $\Delta=4$; we explain why this is 
a general result (for maximal supersymmetry) by elaborating on a comment in \cite{Cowdall:1996tw} concerning the relation of
SS reduction to a generalized ANT construction. This will also allow us to deduce that the possible values for the third 
parameter are independent of which massive deformation of N=8 supergravity we consider.

\subsection{Maximal supergravity and the ANT construction }  

The dimension reduction of 11D supergravity to a $d$-dimensional spacetime produces scalar fields. From the 11-metric we get
scalar fields that form the non-zero entries of an $(11-d)\times (11-d)$ upper-triangular matrix; the diagonal entries are a set of
$(11-d)$ `dilatons' $\boldsymbol{\phi}=(\phi_1, \dots \phi_{11-d})$ while the off-diagonal elements form a set of what we shall refer to,
following \cite{Lu:1995yn}, as  0-form fields: $\{\mathcal{A}_0^{ij}; i<j\}$, for $i,j=1, \dots 11-d$. Another set of 0-form fields $\{A_0^{ijk}; i<j<k\}$ 
comes from the 3-form potential of 11D supergravity. The separation of the scalars into these three sets arises naturally in the construction of 
the $d$-dimensional maximal supergravity by an iterative process in which the Lagrangian for the  dilatons and form fields are found from those
in one higher-dimension. In this process, one finds that the one-form field strengths are 
 \be\label{calF1}
\mathcal{F}_1^{ij} = \gamma^{kj} d\mathcal{A}_0^{ik} \, , \qquad F_1^{ijk} =  \gamma^{li}\gamma^{mj}\gamma^{nk} dA_0^{lmn} \, , 
\ee
where the matrix $\gamma$ is a polynomial function (given in \cite{Lu:1995yn}) of the scalar fields $\mathcal{A}_0$. We pass over
the analogous formulae for non-scalar form-fields as these may all be set to zero, although  the consistency of this truncation for 
$d=4,5$ requires that we also set some scalars to zero (as indicated below). For $d>5$, this truncation of  maximal 
supergravity in the form found in \cite{Lu:1995yn} yields\footnote{We are using a notation in which $|Q_p|^2$  indicates the norm of a $p$-form field $Q_p$, including the usual factor $1/p!$ needed to avoid over-counting due to antisymmetry (and it also includes an inner product in the 7-dimensional space spanned by the dilatons in the case of $|d\boldsymbol{\phi}|^2$.}
\be\label{Lconsist}
L_{d>5}= 2R - |d\boldsymbol{\phi}|^2  - \sum_{i<j<k} e^{{\bf a} _{ijk}\cdot \boldsymbol{\phi}} |F_1^{ijk}|^2
- \sum_{i<j} e^{{\bf b} _{ij}\cdot \boldsymbol{\phi}} |{\cal F}_1^{ij}|^2 \, , 
\ee
where the (11-d)-vectors ${\bf a}_{ijk}$ and ${\bf b}_{ij}$ are given in \cite{Lu:1995yn} (and repeated in \cite{Cowdall:1996tw}). 

The Lagrangian of \eqref{Lconsist} is also a consistent truncation of the maximal supergravity for
$d=4,5,$ provided that we set to zero all but a subset of the 0-form fields $A_0$ for which
\be \label{CSconstraint}
F_1^{[ijk}\wedge F_1^{lmn]} =0\, . 
\ee
An additional special feature of the $d=4$ case is that we may retain the 3-form gauge potential $A_3$ because its field strength ($F_4=dA_3$ after the 
truncation we consider) is the Hodge dual of a scalar. We then have 
\be
L_{d=4} = 2R - |d\boldsymbol{\phi}|^2  - \sum_{i<j<k} e^{{\bf a} _{ijk}\cdot \boldsymbol{\phi}} |F_1^{ijk}|^2
- \sum_{i<j} e^{{\bf b} _{ij}\cdot \boldsymbol{\phi}} |{\cal F}_1^{ij}|^2  -  |F_4|^2  e^{{\bf a}\cdot\boldsymbol{\phi}}\, ,   
\ee
where the 7-vector ${\bf a}$ can also be found in \cite{Lu:1995yn}.  As explained in \cite{Aurilia:1980xj}, 
the $|F_4|^2$ term can be replaced by an equivalent scalar potential term
involving a mass parameter $m$; this yields 
\be\label{4DLag}
L_{d=4} = 2R - |d\boldsymbol{\phi}|^2  - \sum_{i<j<k} e^{{\bf a} _{ijk}\cdot \boldsymbol{\phi}} |F_1^{ijk}|^2
- \sum_{i<j} e^{{\bf b} _{ij}\cdot \boldsymbol{\phi}} |{\cal F}_1^{ij}|^2 -m^2 e^{-{\bf a}\cdot\boldsymbol{\phi}}\, .  
\ee
This is, by construction,  a consistent truncation  of the massive $N=8$ supergravity theory obtained in 
\cite{Aurilia:1980xj}.  We now consider how we may further consistently truncate to the universal one-axion model 
defined by \eqref{basicmodel}. 

The first step is to set to zero all but one of the  $\mathcal{A}_0$ scalars. in which case \cite{Cowdall:1996tw}, 
\be
\gamma^{ij} = \begin{cases}- \mathcal{A}_0^{ij} & i<j\, \\  1 & i=j \\ 0 & i>j\, .\end{cases} 
\ee
If we choose (say) $\mathcal{A}^{12}$ to be the one non-zero scalar (with an 11-metric origin)
then the only off-diagonal component of the matrix $\gamma$ is $\gamma^{12} =- \mathcal{A}_0^{12}$, 
and this does not contribute to the one non-zero one-form field-strengths, which are now 
\be
\mathcal{F}_1^{12}  = d\mathcal{A}_0^{12}\, , \qquad F_1^{ijk} = dA_0^{ijk}\, . 
\ee
This implies the consistency of the truncation since the the non-zero $\mathcal{A}_0^{12}$ can only contribute
to the field equations of the scalars that we set to zero through terms that are manifestly zero when these
scalars are zero. It is also consistent to set to zero any number of the $A_0^{ijk}$ scalars, and we can trivially 
ensure \eqref{CSconstraint} by setting to zero all but one of them, say $A_0^{1'2'3'}$, where 
$\{1'2'3'\}$ is any  order-preserving permutation of $\{123\}$ ({\it i.e.} $1'<2'<3'$).  The 4D Lagrangian of \eqref{4DLag} has now 
been consistently truncated to 
\be\label{4DLagred}
L'_{d=4} = 2R - |d\boldsymbol{\phi}|^2  -  e^{{\bf a} _{1'2'3'}\cdot \boldsymbol{\phi}} |dA_0^{1'2'3'}|^2
- e^{{\bf b} _{12}\cdot \boldsymbol{\phi}} |d\mathcal{A}_0^{12}|^2 -m^2 e^{-{\bf a}\cdot\boldsymbol{\phi}}\, .  
\ee

We now have two ways of arriving a model with a single axion: 
\begin{enumerate} 

\item $A_0^{1'2'3'} =0$ and $\mathcal{A}_0^{12} = \chi$. In this case we can choose, by means of an $SO(7)$ rotation, 
\be
{\bf a} = (-\vec \lambda, 0,0,0,0,0) \, , \qquad {\bf b}_{12} = (\vec\mu, 0,0,0,0,0)\, . 
\ee 

\item $\mathcal{A}_0^{12}=0$ and $A_0^{1'2'3'} =\chi$. In this case we can choose
\be
{\bf a} = (-\vec\lambda, 0,0,0,0,0) \, , \qquad {\bf a}_{1'2'3'} = (\vec \mu, 0,0,0,0,0)\, . 
\ee 
\end{enumerate}
In both cases, we can truncate consistently to the universal one-axion model  by setting 
\be
\boldsymbol{\phi} = (\vec\phi, 0,0,0,0,0) \, , 
\ee
and $(\vec\mu,\vec\lambda)$ are the parameters of this model. We can now use (for $d=4$) the results of \cite{Cowdall:1996tw} for the inner 
products of the $(11-D)$-vector coefficients, to show that (in both cases) 
\be 
\mu=2, \qquad \lambda = \sqrt{7} \quad \Rightarrow \quad  \Delta=4\, , 
\ee
but we find two possible values for $\vec\mu\cdot\vec\lambda$:
\be
(i): \quad \vec\mu\cdot\vec\lambda = - {\bf b}_{12} \cdot {\bf a} = 0\, , \qquad 
(ii): \quad \vec\mu\cdot\vec\lambda = - {\bf a}_{1'2'3'} \cdot {\bf a} = 2\, .
\ee
There was nothing special about the particular choices of axion; either $\chi$ is one of the scalars from the 11D metric, in which case $\vec\mu\cdot\vec\lambda=0$, or it is one of the 0-forms from the 11D 3-form potential, in which case $\vec\mu\cdot\vec\lambda=2$. 

The Lagrangian of \eqref{4DLagred} makes it easy to see why $\mu=2$; it is a direct consequence of the following feature
of the unique {\it massless} maximal supergravity for $d\le9$: 
\be
|{\bf b}_{ij}| = 2\, ,  \qquad |{\bf a}_{ijk}| =2\, , 
\ee
for any (allowed) choice of the indices. It is also true that $\Delta=4$ is a general property of massive maximal supergravities, 
but to see why we first need to consider some more examples.

\subsection{Supersymmetric SS reduction of 11D supergravity}

Toroidal compactification of ten and eleven dimensional supergravity theories lead to effective 
$d$-dimensional supergravity theories, for $d<10$, with many (pseudo)scalar fields; these are massless for `standard' toroidal compactifications but the generalization to flux compactifications\footnote{At least some of the massive supergravity theories discussed in this subsection can be interpreted as flux compactifications; the rest are probably $U$-dual to flux compactifications.} leads to a positive potential for some scalar `dilaton' fields; the remainder we call `axion' fields. The $d$-dimensional supergravity theory can be found by a variant of Scherk-Schwarz dimensional reduction, which introduces a mass parameter \cite{Scherk:1979zr}, but which preserves supersymmetry rather than breaking it \cite{Bergshoeff:1996ui}. 
This method was applied in \cite{Cowdall:1996tw} to toroidal compactifications of 11-dimensional supergravity; for $d=8$ the resulting maximal supergravity Lagrangian can be consistently 
truncated to one in which the only matter fields are (pseudo)scalars: 
\be\label{8dlag}
2e^{-1}\mathcal{L} = 2R - \sum_{i=1}^3 (\partial\phi_i)^2 - 
e^{\vec b_{23}.\vec \phi}(\partial \mathcal{A}_0^{23} )^2
-e^{\vec a_{123}.\vec \phi}(\partial A_0^{123} )^2-m^2 e^{\vec b_{123}.\vec \phi}\, ,  
\ee
where the `dilaton' fields $\{\phi_i; i=1,2,3\}$ are the components of $\vec\phi$, and $\mathcal{A}_0^{23}$, $A_0^{123}$ are 
the `axions'; the notation is that of \cite{Cowdall:1996tw}, and 
we refer the reader to the appendix of that paper for definitions of the three vectors $(\vec b_{123}, \vec b_{23}, \vec a_{123})$, which determine  the dilaton-axion couplings and the scalar potential. Of most relevance here is the fact that these three coupling-constant vectors are linearly independent;  this implies that one cannot eliminate any of the three dilaton fields by means of a consistent truncation unless one of the axion fields is also set to zero. Setting to zero either of the two axions {\it and} the linear combination of dilatons that occurs in its coupling to the three dilatons,  we have a consistent truncation to a one-axion model with a Lagrangian that can be put into the form (\ref{modelogeneral}). In either case, we have 
\be
\vec\lambda = \vec b_{123} \quad \Rightarrow \ \Delta = |\vec b_{123}|^2 - \frac73 = 4\, , 
\ee
but $\vec\mu$ depends on which of the 0-form fields we choose as the axion $\chi$ of the universal one-axion model:
\begin{itemize} 

\item $\chi= \mathcal{A}_0^{23}$. In this case
\be
\mu= |\vec b_{23}| =2 \, , \qquad \vec\mu\cdot\vec\lambda = \vec b_{12} \cdot \vec b_{123} =0\, . 
\ee 

\item $\chi= A_0^{123}$. In this case, 
\be
\mu= |\vec a_{123}| =2\, , \qquad \vec\mu\cdot\vec\lambda = \vec a_{123} \cdot \vec b_{123} =2\, . 
\ee
\end{itemize}
To summarise, we again find that $\mu=2$ and $\Delta=4$, and also that $\vec\mu\cdot\vec\lambda$ is either $0$ or $2$ according to whether the axion has its origin in the 11-dimensional metric or 3-form, respectively. 

Additional $d=7$ maximal massive supergravity theories were found in \cite{Cowdall:1996tw} by SS reduction of the  8D {\it massless} maximal supergravity and, for reasons explained there, they are all different from each other and the $d=7$ maximal massive supergravity found from the 8D massive maximal supergravity that we have just been discussion. In each case there is at least one way to truncate to the universal one-axion model, and one again finds that $\mu=2$, $\Delta=4$, and that 
$0,2$ are the only possible values for $\vec\mu\cdot\vec\lambda$; we pass over the details here because they suggest that the result depends only on the fact that 
our starting point is a maximal supergravity, and this is what we now aim to establish. To do so we make use of a relation between SS dimensional reduction and 
the ANT construction that was mentioned in \cite{Cowdall:1996tw} as a way showing that the massive maximal supergravities obtained by SS reduction of 11D supergravity include the ANT case. The point is that if we first KK reduce to $d=5$ then then we have an $F_4$ field strength that is the Hodge dual of an additional one-form field strength: $\tilde F_1=\star F_4$, where $\tilde F_1 = d\tilde A_0$, for a new axion field. If we KK reduce to $d=4$ there is now no $F_4$ available for 
the ANT construction, but we now have an additional way to SS reduce to $d=4$ by using the shift symmetry of $\tilde A_0$; the resulting 4D massive maximal supergravity theory is the ANT case.

It is possible to reverse this argument, and recover all the massive maximal supergravity theories obtained by SS reduction (at least those that use only a 
single axion and hence have a simple exponential scalar potential) by a generalization of the ANT construction. We now explain this as it provides an easy way
to see why maximal supersymmetry implies a relation between the parameters $(\mu,\lambda)$ of any consistent truncation to the universal one-axion model. 

\subsection{Generalized ANT construction}

We start with a $D$-dimensional theory of gravity coupled to $n$ dilaton fields $\boldsymbol{\psi}  =(\phi_1,\dots,\phi_n)$ and some number of axion fields. 
Choosing one of the axion fields, which we shall call $A_0$, and setting the rest to zero, we have a Lagrangian of the form 
\be\label{LD}
L_D= 2R - |d\boldsymbol{\psi}|^2 - e^{\boldsymbol{\mu} \cdot \boldsymbol{\psi}} |F_1|^2 \, , 
\ee 
where $F_1=dA_0$. We could now perform an SS reduction to $d=D-1$ dimensions, which would generate (after truncation of fields 
that are not scalars in $d$-dimensions) an additional dilaton $\psi_*$  and a positive potential that is an exponential of some linear combination of 
$\boldsymbol{\psi}$ and $\psi_*$. Alternatively, we first dualize the axion field $A_0$ to a $(D-2)$-form potential $A_{(D-2)}$, with $(D-1)$-form field strength 
$F_{(D-1)}= dA_{(D-2)}$, to get 
\be
\tilde L_D = 2R - |d\boldsymbol{\psi}|^2 - e^{-\boldsymbol{\mu}\cdot \boldsymbol{\psi}} |F_{(D-1)}|^2 \, . 
\ee
Next, we KK reduce to $d=D-1$ dimensions, setting to zero the KK 2-form field strength, so the $D$-metric yields the $d$-metric and a scalar, the 
dilaton field $\psi_*$, while from $F_{(D-1)}$ we get  $F_d= dA_{(d-1)}$, which we keep, and $F_{d-1}$ which we set to zero (this is the Hodge dual of 
1-form field strength for the scalar $A_0$ that we get from a direct KK reduction prior to dualization). Using the ansatz
\be
ds^2= e^{-2\beta\psi_* (x)} g^{(d)}_{\mu\nu} dx^\mu dx^\nu+e^{2(d-2)\beta\psi_*(x)}dy^2\, \, , 
\ee
where $\beta$ is defined in \eqref{alphabeta}, 
we then arrive a $d$-dimension theory of gravity with Lagrangian of the form
\be
\tilde L_d = 2R - |d\boldsymbol{\psi}|^2 - |d\psi_*|^2 - e^{-\left[\boldsymbol{\mu} \cdot \boldsymbol{\psi} + 2\alpha \psi_*\right]} |F_{d}|^2 \, ,
\ee
where $\alpha $ is the other constant defined in \eqref{alphabeta}.
The field equation for $A_{(d-1)}$ has the first integral
\be\label{fi}
\star F_d = m\, e^{\boldsymbol{\mu}\cdot\boldsymbol{\psi} + 2 \alpha \psi_*} \, , 
\ee
where $\star F_d$ is the scalar Hodge dual of $F_d$, and $m$ is the integration constant. It is not legitimate to use \eqref{fi} to
eliminate $F_d$ in $\tilde L_d$ when $m\ne0$ because $\tilde L_d$ is not extremized for such solutions of the $A_{(d-1)}$ field equation, 
but this can be remedied by adding to the Lagrangian the total derivative $2m\star F_d$. We may then use \eqref{fi} to eliminate
$F_d$ and the result is the new $m$-dependent Lagrangian
\be\label{mdep}
L_d = 2R - |d\vec\phi|^2  - m^2 e^{\vec\lambda\cdot \vec\phi} \, ,   
\ee
where 
\be
\vec\phi=(\boldsymbol{\psi}, \psi_* ) \, , \qquad \vec\lambda=\left(\boldsymbol{\mu}, 2\alpha \right) \, . 
\ee 
This may be checked by verifying that the field equations coincide with those of $\tilde L_d$ when the solution of the $A_{(d-1)}$ field 
equation is chosen to satisfy \eqref{fi}. We now see that 
\be
\label{forlamb}
|\vec\lambda|^2 = |\boldsymbol{\mu}|^2 + \frac{2(d-1)}{d-2} \, ,  
\ee
and hence 
\be\label{formula}
\Delta= |\boldsymbol{\mu}|^2\, . 
\ee
This formula needs careful interpretation because $\Delta$ is a parameter of $L_D$ whereas 
$|\boldsymbol{\mu}|$ is a parameter of $L_d$. 

\subsection{Implications of maximal supersymmetry}\label{subsec:implications}

We are now in a position to determine the possible parameter values of our universal one-axion model when it occurs as a consistent truncation of {\it any} massive maximal supergravity theory. We have already seen from an examination of the 
4D ANT case that two possibilities are 
\begin{eqnarray}\label{pvals}
(i): & \Delta= 4\, , \quad   \mu =2\, ,  \quad  & X =0\, ,   \nonumber \\
(ii): & \Delta= 4\, ,  \quad  \mu =2\, , \quad    & X =\tfrac12\, . 
\end{eqnarray}
A detailed investigation of several different $d>8,7$ supergravity theories led to exactly the same possibilities.
This suggests, and we shall now prove,  that the above two possibilities are the only ones permitted by
maximal supersymmetry.  

If $L_D$ of \eqref{LD} is a truncation of the $D$-dimensional maximal supergravity Lagrangian then $L_d$ of \eqref{mdep} is a truncation of the SS-reduced supergravity theory, and also of any universal one-axion model contained in it; in this case $|\boldsymbol{\mu}|=2$. The formula \eqref{formula} then tells us that $\Delta=4$, both for the one-axion model and the full supergravity theory that contains it. This remains true for any KK reduction, e.g. to 4D, so the Lagrangian of \eqref{4DLag} which was initially applicable only to the ANT model, is also applicable to any of the other massive maximal supergravity theories obtained by SS reduction. This is because $\Delta=4$ enables us to rotate the dilaton vector $\vec\phi$ to bring the vector $\vec\lambda$ in \eqref{mdep} to the `cosmological' dilaton vector of \eqref{4DLag}. As the parameters of the one-axion model obtained by  consistent truncation of this Lagrangian are not affected by this rotation, we conclude that 
the possible sets of these parameters are the same for all massive maximal supergravity theories obtainable by SS reduction of 11D supergravity, and that these are the values found for the ANT case. 

The only case omitted by this argument is the maximal massive 4D supergravity obtained by dimensional reduction of the 10D mIIA supergravity theory (equivalently, by SS reduction of the 10D IIB supergravity), but in this case it is known that $\Delta=4$, and this is unaffected by KK reduction to 4D and subsequent consistent truncation to  \eqref{4DLag}, at which point the logic applied above to SS reduction from 11D supergravity again applies. 
The two sets of values given in \eqref{pvals} are therefore the only possibilities for the universal one-axion model if we require it to be a consistent truncation of a maximal 4D supergravity theory.  

Furthermore, the fact that there are two possibilities for $\vec\mu\cdot\vec\lambda$ is explained by the fact that 
there are two possible higher-dimensional origins for the axion of the universal one-axion model embedded in any maximal massive supergravity theory. For those cases with an 11D supergravity origin, the axion can come either from the
11D metric (in which case $\vec\mu\cdot\vec\lambda =0$) or from the 11D 3-form  potential (in which case $\vec\mu\cdot\vec\lambda =2$). 

Now that we know the possible values of the parameters appearing in the universal one-axion model in the current context, we can apply the results of section \ref{sec:FLRW}. Recall that fixed points of the three-dimensional autonomous dynamical system can exist only if $A>0$.
This condition is equivalent to $\Delta <\mu^2 X^2$, which implies $X^2>1$ for $\Delta=4$, but this is not satisfied either by $X=0$ or by $X=\tfrac12$.
There are therefore no scaling solutions at fixed points of the autonomous dynamical system \eqref{3Dsystem}. All trajectories must start and end on the circle $x^2+y^2=1,\ z=0$. The  $X =0$ case corresponds to $\theta =\frac{\pi}{2}$, which is the special case discussed at the end of section \ref{sec:FLRW}. The $X= \tfrac12$ case is qualitatively similar since trajectories again start and end on the fixed-point circle $x^2+y^2=1, \ z=0$, with the endpoint lying in the attractive part of this circle  (depicted in green color in the figures); the trajectories are qualitatively as in fig. \ref{primera}. As these trajectories cover the full phase space, some interior trajectories pass very close to the region $z=1$, where $w\approx -1$; in this region there is a transient de Sitter-like phase, but this phase occurs only once. There is also one boundary trajectory that passes through $z=1$;  on this trajectory the scalar kinetic energy is zero when the value of the potential is at its maximum.

\section{Summary and discussion}

Motivated by the structure of ``massive'' deformations of maximal supergravity theories in $d\le8$ 
spacetime dimensions, such as those found from dimensional reduction of the Romans d=10 mIIA supergravity, 
but including many others found by  a supersymmetry-preserving Scherk-Schwarz dimensional reduction of 
$d=11$ supergravity,  we have introduced a simple three-parameter model of an axion field coupled to two dilaton fields, all coupled to gravity. By requiring invariance under a constant rescaling of the metric combined with a constant shift of one linear combination of the two dilaton fields, the scalar field interactions are restricted to an exponential dilaton potential (with gradient $\vec\lambda$) and an exponential dilaton coupling to the axion field (with gradient $\vec\mu$); the three parameters are the magnitudes of the 2-vectors $(\vec\lambda,\vec\mu)$
and the angle $\theta$ between them.

Another motivation for this model is that it generalizes the two-parameter dilaton-axion model with a single dilaton  that was shown in  \cite{Sonner:2006yn} to have flat FLRW cosmologies that undergo recurrent periods of 
cosmic acceleration. If this phenomenon is possible in string/M-related supergravity theories 
it could help resolve the tension between string/M-theory and $\Lambda$CDM cosmology caused by the  difficulty/impossibility of finding de Sitter compactifications. However, the one-dilaton model of \cite{Sonner:2006yn} is not a consistent truncation of any maximal supergravity theory; this is one 
of our results here, but we also find that the two-dilaton model can be found by a consistent truncation in two different ways. 

A notable feature of our two-dilaton/one-axion model, which led us to refer to it as the ``universal one-axion'' model, is that it reproduces itself under a process of dimensional reduction/truncation but with a change of parameters. Furthermore, by considering the spacetime dimension $d$ as another (discrete) parameter, we can 
replace the three continuous parameters for given $d$ with an alternative set of three parameters $(\Delta,\mu,X)$ that are both $d$-independent and independent of the $O(2)$ basis chosen for the two dilatons.  Because of this dimension independence, any consistent truncation of a $d$-dimensional supergravity model to the universal one-axion model implies a similar  consistent truncation of the supergravity theory reduced to $d=4$, with the same three parameters. 

It would be possible to define analogous ``universal $n$-axion'' models with $(n+1)$ dilatons, in which case
more complicated potentials would be possible. We expect that some such models for $n>1$ could be found 
as consistent truncations of massive maximal supergravity theories with multiple mass parameters as these
have potentials that are sums of exponentials of different linear combinations of dilatons. This may be
a fruitful avenue for further research, but we expect that it will be difficult to acquire any detailed understanding of cosmological solutions for $n>1$. This is another reason for our focus on the one-axion case, but there is a also a physics justification:  in the context of FLRW cosmology driven by scalar fields with a positive sum-of-exponentials  potential, it is expected that a single exponential will dominate at late times. 

Another notable feature of the universal one-axion model is that a study of its flat FLRW cosmological solutions reduces to finding trajectories of a three-dimensional autonomous dynamical system, as a function of the three parameters $(\lambda,\mu,\theta)$; this generalises the two-dimensional system analysed in \cite{Sonner:2006yn}.  Additionally, a complete qualitative understanding of the trajectories of this system is both possible and fairly  simple; this was not guaranteed {\it a priori} as some superficially-similar well-known three-dimensional autonomous systems are chaotic. For some ranges of parameters, the autonomous system has fixed points
corresponding to scaling solutions of the Friedmann and scalar field equations. The possibility of eternal cosmic acceleration occurs at these fixed points, and we have found the necessary and sufficient conditions on parameters that allow this possibility, or exclude it. 

For sufficiently weak axion-dilaton coupling ($\mu<\mu_c$) eternal cosmic acceleration is excluded if the dilaton self-coupling is sufficiently large ($\lambda := |\nabla V|/V \ge \lambda_c$). The similarity of this condition to the cosmological swampland bound on $\nabla V$ that has been proposed and refined in recent years suggests that these bounds are consequences of a principle of no eternal acceleration; this is plausible because eternally accelerating FLRW spacetimes have a future cosmological event horizon, which is a known impediment to the construction of a consistent theory of quantum gravity.  On this hypothesis, we have interpreted the no-eternal-acceleration condition on the parameters of the universal one-axion model as swampland bounds. In addition to making precise the swampland bound on $\nabla V$,  we find that for strong axion-dilaton coupling there is a stronger swampland bound that can be 
viewed as an upper bound on the curvature of the scalar-field target space. We are not aware of any previous discussion of swampland bounds of this type; if future work shows that they arise naturally from string/M-theory then this could support the hypothesis that cosmological swampland bounds are consequences  of a quantum gravity prohibition of future cosmological event horizons. 

Although the no-eternal-acceleration principle rules out eternal cosmic acceleration, it does not rule out the possibility of transient cosmic acceleration, and for a large range of parameters this will occur for a large 
range of initial conditions. However, of most interest for potential phenomenological applications are those
cosmological trajectories for which the spacetime most resembles de Sitter space during the transient acceleration phase, or phases in the case of recurrent transient acceleration. We have quantified this requirement by 
introducing an equation of state in which the pressure to density ratio $w$ evolves with cosmological time; 
in this context trajectories on which $w\approx -1$ during some transient acceleration phase may be said to be 
de Sitter-like to the same approximation. 

In our application of the universal one-axion model to maximal supergravities, neither of the two 
possible sets of  parameters allows eternal acceleration; there are cosmological solutions that pass through 
a phase of de Sitter-like expansion, but only once since recurrent transient cosmic acceleration 
is also not possible for the parameters permitted by maximal supersymmetry. However, we expect that 
other parameter ranges of the universal one-axion model will 
occur in consistent truncations of other (non-maximally supersymmetric) compactifications of string/M-theory, 
allowing a richer set of cosmological possibilities.


\section*{Acknowledgments}

JGR acknowledges financial support from project  MINECO grant PID2019-105614GBC21.
PKT is partially supported by the STFC consolidated grant ST/T000694/1


\end{document}